\documentclass[twocolumn,times,trackchanges]{aastex631}

\usepackage{graphicx}	
\usepackage{amsmath}	
\usepackage{bbding}
\usepackage{hyperref}

\shorttitle{Larger-Scale Cold-Gas AM Environment and Galaxy SF}
\shortauthors{Wang et al.}

\begin{document}


\title[]{From larger-scale cold-gas angular-momentum environment to galaxy star-formation activeness}

\author[0009-0007-6358-3564]{Sen Wang}
\affiliation{Department of Astronomy, Tsinghua University, Beijing, 100084, China}
\correspondingauthor{Sen Wang}
\email{wangsen19@mails.tsinghua.edu.cn}

\author{Dandan Xu}
\affiliation{Department of Astronomy, Tsinghua University, Beijing, 100084, China}
\correspondingauthor{Dandan Xu}
\email{dandanxu@tsinghua.edu.cn}

\author[0000-0002-6726-9499]{Shengdong Lu}
\affiliation{Institute for Computational Cosmology, Department of Physics, University of Durham, South Road, Durham, DH1 3LE, UK}






\begin{abstract}

We study the influence of the ambient large-scale cold-gas vorticity on the specific star formation rate (sSFR) of central galaxies with stellar masses of $10.0<\log\,M_{\ast}/\mathrm{M_{\odot}}<11.5$ at $z=0$, using the TNG100 simulation. The cold-gas vorticity defined and calculated for gas with $T_{\rm gas} < 2\times 10^4 \mathrm{K}$ and on scales of $\sim$ 1 Mpc can well describe the angular motion of the ambient cold gas. We find crucial evidence for connections between the cold-gas vorticity and star-formation activeness, such that at any given halo mass (particularly below $10^{13}\,\mathrm{M_{\odot}}$), galaxies living in higher cold-gas vorticity environments are generally less star-forming, regardless of their large-scale environment types (filament or knot), 
or star formation states (star-forming or quenched). Specifically, at a fixed halo mass scale of $10^{12}-10^{13}\,\mathrm{M_{\odot}}$, the median sSFR of galaxies living in environments with the top 30\% cold-gas vorticity is $\sim 0.5$ dex below that of galaxies living in environments with the bottom 30\% cold-gas vorticity. At any fixed halo mass scale, cold-gas vorticities around filament galaxies are generally higher than those around knot galaxies, consistent with that filament galaxies have lower sSFRs than knot galaxies. This large-scale cold-gas vorticity is highly connected to the orbital angular momentum of neighboring galaxies up to a distance of $\sim$ 500 kpcs, indicating their common origin and a possible angular momentum inheritance/modulation from the latter to the former. The negative modulation by the environmental vorticity to galaxy star formation is only significantly observed for the cold gas, indicating the unique role of cold-gas angular momentum. 


\end{abstract}



\section{Introduction} 

\label{sec:introduction}

The angular momentum environment of galaxies plays an important role in shaping and regulating a variety of galaxy properties, including their morphologies, kinematics, as well as star formation activities (e.g., \citealt{Fall_Efstathiou_1980, Mo_Mao_White_1998, Teklu_2015, Zavala_2016_AMEagle, Rodriguez-Gomez2017, ZengWang_2021MergerMassiveDisk, Lu_2021_HotSFDisk, Wang_et_al_2022, Lu_et_al_2022, Lu_2022_ColdQuenched, Valenzuela_2024_MergerSpin, HuXuLi2024RAASuperthin, Gamez-Marin24_Satellite_CosmicWeb}). Such angular momentum environments can refer to a wide range of angular momentum properties, for example, the theoretical halo and galaxy spins (e.g., \citealt{Peebles_1969, Mo_Mao_White_1998, Bullock_2001, Bett_2007_HaloSpin}), the observed stellar angular momentum (e.g., \citealt{Emsellem_2011_StellarAMFRSR, Romanowsky_2012, Cortese_2016}), the in-spiral motion of the circumgalactic gas (e.g., \citealt{Stewart_2011, Ho_2017_CGMAcc,Zhang_et_al_2023}), the orbital angular momentum of neighbouring galaxies at distances up to sub-Mpc scales (\citealt{Libeskind_2012, Libeskind_2014_SubhaloAccAndVF, Danovich_2015_CosmicWebStreamAM, Lu_et_al_2022, Wang_et_al_2022, Wang_2024_SAMIMaNGA, Valenzuela_2024_MergerSpin}), as well as the cosmic vortical flow and filamentary accretion on Mpc scales (e.g., \citealt{PinchonBernardeau99, Pichon_2011_FilamentRichAM, Libeskind_et_al_2013_a, Libeskind_et_al_2013_b, Danovich_2012_CoplanarStreamsAM, Stewart_2013_FilamentAccAndMergerAM, Hahn_2015_CosmicVelocityField, Laigle_2015_FilaVorticesDMhaloSpin, Lu_2024_FilamentAM}). We have come a long way understanding the many processes and mechanisms related to the formation and evolution of these angular momentum properties.

In the current paradigm, the initial angular momentum of a dark matter halo is not the result of primordial velocity {\it vector} perturbation, which would soon decay due to the cosmic expansion (\citealt{Jones_1976, Gott_1977, Efstathiou_Silk_1983, Bernardeau_2002}), but a consequence of a torque moment exerted by a shear, curl-free, potential flow of the ambient large-scale structure in the linear or quasi-linear regime prior to the gravitational collapse of the region, as is explained by the tidal torque theory (e.g., \citealt{Peebles_1969, doroshkevich1970, Fall_Efstathiou_1980, white1984, barnes1987, Catelan_Theuns_1996, Schaefer_2009_Review}, and also see \citealt{LiaoSH_2017, Neyrinck_2020}). In an isolated scenario, this angular momentum is conserved during the collapse after turnaround. Gas condenses within the dark matter halo, forming gaseous discs where star formation happens. The newly formed stellar disc is therefore expected to retain a large fraction of the specific angular momentum of the dark matter halo (e.g., \citealt{WhiteRees_1978, Mo_Mao_White_1998, Firmani_Avila-Reese_2000, Firmani_Avila-Reese_2009, RAR_2023}). 

In reality, however, galaxies live and flow in the cosmic web. Their angular momentum acquisition, in particular that is associated with the cold stream accretion in filamentary structures, can also be influenced by the rich dynamics due to a high level of non-linearity of the cosmic web on large scales. This effect can be particularly strong at high redshifts (\citealt{Pichon_2011_FilamentRichAM, Danovich_2012_CoplanarStreamsAM, 
Codis_2012_HaloSpinFilament, Libeskind_et_al_2013_b, Danovich_2015_CosmicWebStreamAM, Laigle_2015_FilaVorticesDMhaloSpin, Tillson_2015_HighZAMFromColdAccFilament, Lu_2024_FilamentAM}). In addition, galaxies in their dark matter halos also strongly interact and merge among themselves to grow more massive structures in a hierarchical fashion. These highly non-linear processes of galaxy mergers and flybys also largely influence the growth of angular momentum (e.g., \citealt{Vitvitska_2002_MergerHaloSpin, 
Maller_DS_2002_MergerHaloSpin, 
Hetznecker_2006_MinorMergerUpAM, Bois_2011_MergerFRSRSpin, Welker_2014_MergerHaloSpin, BettFrenk_2016}). These processes can not only change the angular momenta of dark matter halo and stellar discs, but also inject angular momentum into the circumgalactic medium (CGM), in particular the cold CGM gases that are accreted to galaxies (e.g., \citealt{Stewart_2011, Stewart_2013_FilamentAccAndMergerAM, Wang_et_al_2022, Lu_et_al_2022}).   
We shall also note that apart from the variety of dynamical effects due to gravitation as discussed above, kinetic and thermal feedback of baryons can also change a galaxy's angular momentum environment (e.g., \citealt{Maller_Dekel_2002_FeedbackHaloSpin, Bett_2010_BaryonHaloSpin, Sharma_2012_HaloMerger, Genel_2015_FeedbackSpin, DeFelippis_2017_FeedbackSpin}) by altering the angular momentum distribution of any given matter component or transferring angular momenta among different matter components (see e.g., \citealt{Zavala_2016_AMEagle, zjupa2017}).

In this study, we focus on a specific type of angular momentum modulation that is closely related to the galaxy interaction environment and the large-scale vortical flow. In particular, we study the preventative impact of this angular momentum environment on galaxy star formation activities. We note the reader that \citet{Peng&Renzini} proposed the idea of excessive angular momentum inhibiting disk star formation based on observed scaling relations. \citet{FilamentEdgeSong2021} using the HORIZON-AGN simulation (\citealt{HorizonAGN2014D}) revealed a suppression of star formation near the edge of filaments where the gas vorticity is higher than at the centre (see the two bottom left panels in figure 3 therein). The authors also suggested that gas transfer to galaxies become less efficient due to high angular momentum supply in the vorticity-rich edge of filaments, and further proposed that this may act as a quenching mechanism for a large fraction of passive galaxies in filaments.

Following up on a recent paper series (\citealt{Wang_et_al_2022, Lu_et_al_2022}), we investigated the relation between environmental angular momenta and galaxy star formation activities using typical star-forming and quenched galaxy samples from the TNG100 simulation (\citealt{Pillepich_et_al_2018b, Marinacci_et_al_2018, Springel_et_al_2018, Naiman_et_al_2018, Nelson_et_al_2018, Nelson_et_al_2019a}). We found that the CGM and the ambient galaxy environment around present-day quenched galaxies tend to have higher spins and higher orbital angular momenta than those of star-forming galaxies (within a similar mass range). Our interpretation of that finding was also in line with the above-mentioned studies and that a higher CGM spin, as inherited from interacting galaxies in the neighborhood, can prevent efficient gas infall to fuel central star formation. While for the massive red and dead galaxies, this provides an alternative mechanism that can keep them on lower star formation activities (once it is quenched), in particular being complementary to the AGN feedback quenching mechanism (e.g., \citealt{Granato_et_al.(2004),Hopkins_et_al.(2005),Di_Matteo_et_al.(2005),Di_Matteo_et_al.(2008),Croton_et_al.(2006),Cattaneo_et_al.(2009)}).

In this study, we aim at searching for stronger evidence using the simulation for a close connection between a galaxy's star formation activeness and its ambient angular momentum environment among present-day galaxies. Regarding our previous studies, there are three aspects worth noting. Firstly, both quenching and halo spin correlate positively with galaxy/halo mass, and the observed connection between spin and star formation rate (SFR) could merely be a consequence of such mass dependencies, although in \citet{Lu_et_al_2022} we had adopted a few different definitions of spin in order to eliminate the mass dependence. Secondly, it is the cold gas (e.g., with an effective temperature $T_{\rm gas}$ below $\sim 2\times 10^4$\,K, see Section 2 for details) that is directly relevant for star formation. Therefore, the spin property in question is neither that of the entire halo, nor that of the total gas, but only of the cold gas. Thirdly, from a gas acquisition perspective, the cold circumgalactic gas at larger scales is often highly localized in the form of extended streams, sometimes accreted from the intergalactic medium associated with filamentary structures, and sometimes stripped from incoming satellite galaxies (e.g., see figure 7 and 8 in \citealt{Wang_et_al_2022} and the related discussion therein). Therefore conventionally defined spin parameters (e.g., \citealt{Bullock_2001}) are not suited to capture such angular momentum behaviors. 

Due to these reasons, we need to find a better definition for ``spin'' that can describe the angular motion of the ambient cold gas at larger distances, which shall also not necessarily be confined to a halo’s domain. We therefore adopt the velocity vorticity $\nabla \times \boldsymbol{v}$ for the cold gas on a spatial scale of $\sim 1$\,Mpc (see Section 2 for details). We note that the vorticity field of the cosmic matter flow is a result of higher-order perturbation developed in the highly non-linear regime. It has been well studied and found to be largely responsible for halo angular momentum growth during the non-linear evolution phase at high redshifts (e.g., \citealt{PinchonBernardeau99,  Bernardeau_2002,  Wang_AC_2014_RotationalFlow}), and also to explain the alignment between halo spins and the large scale structures (e.g., \citealt{Libeskind_et_al_2013_b, Hahn_2015_CosmicVelocityField, Laigle_2015_FilaVorticesDMhaloSpin}). In this study, we also investigate this quantity but for the ambient cold gas at larger scales, and take this as a more general depiction of its angular motion. While the choice of 1\,Mpc is motivated by our previous studies (\citealt{Wang_et_al_2022, Lu_et_al_2022}) which have demonstrated that the cold circumgalactic gas (with an in-spiraling streaming motion at distances of $\sim$300\,kpc scales), as influenced by the angular momentum environment on larger scales, can substantially regulate the efficiency of the cold gas infall from that large distances. In many occasions, such an angular momentum modulation, is closely related to galaxy/subhalo interaction and merger that happen at that large distances, which help to enhance the CGM angular momentum in question.

With this, we can now, at a more fundamental level, search for clear evidence for connections between the cold-gas spin/vorticity and star formation activeness. Specifically, we switch the perspective from specific angular momentum of CGM up to $\sim$300\,kpc and gravitationally bounded to galaxy halos as in our previous studies (\citealt{Wang_et_al_2022, Lu_et_al_2022}), to the cold-gas vorticity of the cosmic environment on $\sim 1$\,Mpc scale as in this study. We find that at any given mass, galaxies that live in a higher cold-gas vorticity environment are generally less actively star-forming. We study how this cold gas vorticity may vary in different large-scale cosmic environments and find that at any fixed halo mass scale, cold-gas vorticities around filament galaxies are generally higher than those around knot galaxies. It is interesting to note that this happens to go in line with the fact that galaxies living in filament also have lower sSFRs than knot galaxies (due to their slower assembly process and lower-density environment). We also demonstrate that this quantity, as expected, is highly connected to the orbital motions of galaxies in the vicinity (also see \citealt{Libeskind_2014_SubhaloAccAndVF}), indicating their common origins and a possible angular momentum inheritance/modulation from the latter to the former. We show that such a negative modulation by the angular momentum environment to galaxy star formation is only significantly observed for the cold gas, but neither for the total matter nor the total gas component, indicating the unique role of cold gas. We further emphasize that the angular/vortical motion of the cold gas on large scales, is a crucial element shaping the star-forming and quenching status of a galaxy; the cold-gas vorticity field is thus an important feature, among others, to depict a galaxy’s larger-scale environment when studying the dependence of galaxy evolution.

This paper is organized as follows: In Section \ref{sec:method}, we present the simulation details and illustrate the on-mesh definitions of a number of properties and the method that we use to classify large-scale structures. In Section \ref{sec:old}, we re-visit our typical galaxy samples from previous works and emphasize the uniqueness of cold gas. We then study the cold gas vorticity environment of general galaxy samples and samples with different star formation states and environment types in Section \ref{sec:general_sample}. Finally discussion and conclusions are given in Section \ref{sec:discussion} and \ref{sec:conclusion}, respectively. Throughout the paper, we adopt the same cosmology as those used in the IllustrisTNG simulation (i.e., based on the Planck results \citealt{Planck_Collaboration(2016)}), with a total matter density of $\Omega_{\rm m} = 0.3089$, a baryonic matter density of $\Omega_{\rm b} = 0.0486$, and a Hubble constant $h = H_0/(100\,{\rm km s}^{-1} {\rm Mpc^{-1}}) = 0.6774$, assuming a flat $\Lambda$CDM universe.

\section{Methodology} 

\label{sec:method}

\subsection{The simulation}

\textit{The Next Generation Illustris Simulations} (IllustrisTNG, TNG hereafter; \citealt{Marinacci_et_al_2018, Naiman_et_al_2018, Springel_et_al_2018, Pillepich_et_al_2018b, Nelson_et_al_2018,  Nelson_et_al_2019a}) are a suite of magneto-hydrodynamic cosmological simulations implemented by the moving-mesh code \textsc{arepo} \citep{Springel_2010} for galaxy formation and evolution. The simulation includes a subgrid model with sub-resolution ISM stochastically forming stars with a density threshold of $\sim 0.13\,{\rm cm}^{-3}$, radiative cooling, stellar evolution and chemical enrichment, stellar and AGN feedback. Detailed physical model prescriptions can be found in \citet{Vogelsberger2013, Pillepich2018a, Weinberger2017}. In this paper, following the previous works, we use the TNG100 simulation, which carries a periodic box of side length of 110.7 Mpc with mass resolutions of $7.5\times10^6\,{\rm M_{\odot}}$ and $1.4\times10^6\,{\rm M_{\odot}}$ for the dark matter (DM) and the baryon. The gravitational softening length is $0.5\,\mathrm{h^{-1}kpc}$ for the dark matter and stellar particles. The host halos are identified using {\sc subfind} algorithm \citep{Springel_et_al_2001,Dolag_et_al_2009}. The outputs of galaxies and halos of all 100 snapshots are available through the public data access \citep{Nelson_et_al_2019a} of TNG project\footnote{\url{http://www.tng-project.org/data/}}.

\subsection{Basic setup of grid} 

The goal of this study is to identify and understand the correlation between the large-scale angular momentum field in the form of the flow vorticity on a scale of $\sim 1$\,Mpc 
and the galaxy star formation activeness. The choice of this distance range is motivated by previous studies. In \citet{Wang_et_al_2022} and \citet{Lu_et_al_2022} we investigated the impact of galaxy interaction environment in terms of orbital angular momenta in galaxy vicinities out to a radius of 300 kpc, which is roughly the virial radius of dark matter halos just above the Milky-Way size (e.g., \citealt{Aquarius2008S, Grand_et_al_2017}). We found that neighboring galaxies out to at least such distances transfer their orbital angular momenta to the cold circumgalactic gas of the host halos through galaxy interactions like megers, fly-bys, tidal stripping, and so on. Such angular momenta are then further transported to the stellar discs where newly-formed stars inherit the angular momentum of the accreted cold gas. It is worth noting that the TNG100 galaxies already demonstrated coherent kinematics among the host galaxy (stellar) discs, the cold CGM gases and the galaxies in the neighborhoods (see \citealt{Lu_et_al_2022}). Observationally, such a coherent motion has been reported by \citet{Wang_2024_SAMIMaNGA} among the MaNGA (\citealt{Bundy_et_al_2015}) and SAMI (\citealt{Croom_et_al_2012}) galaxies and their satellite galaxies out to $\sim$ 100 kpc in their vicinities. While \citet{Lee_et_al_2019a} used the CALIFA (\citealt{Sanchez_et_al_2012}) survey data and found a significant dynamical coherence between galaxy rotation and the average motion of neighbor galaxies out to distances of $\sim$ 800 kpc.

For this reason, we have implemented a grid mesh of $128^3$ to cover the entire simulation box of (110.7 Mpc)$^3$ with spatial resolution of $\sim 865$ physical kpc. This is our basic mesh setup, which is used to calculate several fields, including (1) specific star formation rate (sSFR, as contributed by all central galaxies), (2) total matter over-density, (3) velocity vorticity, and (4) large-scale environmental type. We note that the grid cells themselves are also used as statistical samples when comparing among the total matter vorticity, the total gas vorticity and the cold-gas vorticity (see Fig.\,\ref{fig:Vorticity_Overdensity}). 

For each grid cell, we first compute the sSFR on the grid. Specifically, we choose all central galaxies at redshift $z=0$ (snapshot 099) and add their stellar masses and SFRs to the grid using the near-grid point assignment. The ratio of the two gives the sSFR of the grid cell. We also calculate the total matter over-density field, defined as $\delta \equiv \rho_{\rm m} / \bar{\rho}_{\rm m} -1 = \rho_{\rm m} / (\Omega_{\rm m} \rho_{\rm cr}) -1$, where $\rho_{\rm cr}$ is the critical density of the universe. For this, we take all particles at redshift $z=0$ and assign them to the grid according to the {\it Clouds in Cells} (CIC) algorithm.

\subsection{Calculating vorticity fields on the grid}

In previous our studies \citep{Wang_et_al_2022,Lu_et_al_2022}, we demonstrated an environmental modulation from the galaxy neighborhood (satellite orbital motion) to the CGM spin (see figure 4 in \citealt{Lu_et_al_2022}) and further to the central sSFR (see figure 5 in \citealt{Wang_et_al_2022} and figure 2,\,3 in \citealt{Lu_et_al_2022}). Here in this work, we would like to generalize the CGM angular momentum property to the vorticity of velocity fields, a similar concept to the cosmic vortical flow on large scales (\citealt{PinchonBernardeau99, Wang_AC_2014_RotationalFlow, Libeskind_et_al_2013_b, Hahn_2015_CosmicVelocityField}).

To do so, we compute the vorticity field $\boldsymbol{\omega}$ for three types of components, i.e., total matter, total gas, and the cold gas. Each time we take all either particles, or total gas cells, or the cold-gas cells alone, and assign them to the grid to obtain the corresponding velocity field according to the CIC algorithm without further smoothing procedures. In addition, we have also calculated the velocity fields using the {\it Smoothed particle hydrodynamics} (SPH) algorithm (\citealt{Monaghan1992}) with smoothing length $h$ given by the comoving radius of the sphere enclosing 64 nearest dark matter particles. We have verified that all of our final results remain unchanged.

According to the Helmholtz Decomposition Theorem, one can extract the rotational parts of a general vector field by applying a curl operator to the field. Considering the kinetic motion of a region, this ``rotation'' can be directly defined as the curl of the velocity field $\boldsymbol{v}$, namely the vorticity field $\boldsymbol{\omega}\equiv \nabla \times \boldsymbol{v}$. Once we have obtained a velocity field, we calculate the curl of this field using the gradient function in the {\sc python} package {\sc NumPy}\footnote{\url{https://numpy.org}}. The vorticity field $\boldsymbol{\omega}$ calculated on the mesh essentially reveals the surrounding (grid scale) ``rotational'' motion relative to the target grid point. We note that in this study we mainly focus on the norm of the vorticity field, i.e. the larger or smaller of the vorticity in a galaxy's (cold-gas) environment, and its relation to galaxy star formation activeness. It is worth nothing that the direction of the vorticity field at each grid point is also important and interesting to study because it indicates the path of matter accretion in relation to the large-scale structure of the cosmic web. We defer this to a further study.

\subsection{Cold gas}

Specifically, we follow \citet{Wang_et_al_2022} and define the cold gas using the effective temperature threshold $T_{\rm gas} < 2\times 10^4 \mathrm{K}$. 
We note that around and below $2\times 10^4\, \mathrm{K}$ is a typical temperature range of atomic hydrogen, which may cool to form stars inside galaxies either directly assembling central disks in less massive halos or as cold streams penetrating through the shock-heated medium in more massive systems (\citealt{Keres2005, Dekel2006, Dekel2009Nature}). Observationally, this temperature range is also largely dominated by atomic hydrogen in a warm and diffuse phase, contributing to Ly-alpha line (\citealt{KatzGunn1991, Fardal2001ColdGasLyA}) and HI 21 cm emission (\citealt{Draine2011book, SaintongeCatinella2022}). We note that in TNG100 the total mass fraction of the gas in this temperature range in the entire simulation box is 27.5\% at redshift $z=0$.  

In this study, we focus on the influence of the angular momentum environment particularly from the diffuse cold gas on larger scales, instead of those in the close vicinity to any galaxies. This is because the central gas motion as well as its relation with star formation activeness can be largely affected by feedback processes, which are implemented differently among different simulations. To avoid of such contamination, we therefore have also excluded all the cold gas elements that are gravitationally bound within 1/3 of any galaxy's $r_{200}$ (which is the radius within which the mean mass density is 200 times the critical density of the universe). We note that within each galaxy halo, averaged mass fractions of the gas within this temperature range are 66.4\% and 26.1\%, respectively, for those located within $2\,R_{\rm hsm}$ (where $R_{\rm hsm}$ is the half stellar-mass radius of a galaxy) and those outside $r_{200}/3$. We have verified that all results calculated in the latter case remain the same qualitatively as using all cold gas elements. For this reason, in all figures below, we present results using all cold gas elements in the simulation, except for Fig.\,\ref{fig:ssfr_Vorticity_Binned} where both cases are presented for comparisons as a demonstration.   

\subsection{Cosmic web construction and classification}
\label{sec:method_cosmicweb}
In order to understand the cold-gas vorticity modulation in different large-scale structure environment, we have also classified the cosmic web structure of the simulation into four basic types, i.e., knot, filament, sheet, and void. There have been many methods developed for such classification schemes, for example, using the Hessian of the gravitational potential \citep{Hahn_et_al_2007,Romero_et_al_2009}, the shear of the velocity field \citep{Hoffman_et_al_2012,Libeskind_2012,Libeskind_et_al_2013_a,Libeskind_et_al_2013_b,Libeskind_et_al_2014}, or based on topology of the density field, e.g., DisPerSE \citep{Sousbie_2011, Sousbie_et_al_2011}, and so on. In this study, we follow the approach developed by \citet{Hahn_et_al_2007} and \citet{Romero_et_al_2009}, which is a dynamical classification scheme of the large scale environment, based on the gravitational potential field of the total matter distribution. Here we present a summary of our implementation of this method. 

To characterize the cosmic web and the surrounding large-scale structure of galaxies, we adopt the same mesh setup as we use for the vorticity field calculation (see Section~\ref{sec:method_cosmicweb}); in particular, the mesh resolution is compatible to those used in \citet{Hahn_et_al_2007, Romero_et_al_2009}. We first take all particles and assign them to the mesh according to the CIC scheme, yielding the total matter density field $\rho_{\rm m}$.

The potential field $\phi_{\rm m}$ can then be derived by solving the Poisson's equation from the matter density field via:
\begin{equation}
\label{Poisson}
\nabla^2\phi_{\rm m} = 4\pi G \rho_{\rm m}.
\end{equation}
This calculation can be achieved via the Fast Fourier Transform technique. A deformation tensor $T_{\rm ij}$ is then defined as the Hessian of the gravitational potential: 
\begin{equation}
\label{Hessian}
T_{\rm ij}\equiv \frac{\partial^2 \phi_{\rm m}}{\partial x_{\rm i}\partial x_{\rm j}}.
\end{equation}
\label{sec:Web}
At each mesh grid point, $T_{\rm ij}$ is again calculated using the gradient function with second order accuracy in the {\sc python} package {\sc NumPy}.

The eigenvalues of $T_{\rm ij}$, denoted as $\lambda_{1},\,\lambda_{2},\,\lambda_{3}$ (with $\lambda_{1}\geq\lambda_{2}\geq\lambda_{3}$), are then calculated for all grid points. In \citet{Hahn_et_al_2007}, having adopted a threshold value of $\lambda_{\rm th}=0$, the authors classified a grid point as knot, filament, sheet, or void according to the number of eigenvalues above $\lambda_{\rm th}$. With $\lambda_k > \lambda_{\rm th}$ ($\lambda_k < \lambda_{\rm th}$) implying collapse (expansion) along the $k$th eigenvector, the number of eigenvalues above $\lambda_{\rm th}$ essentially corresponds to the dimension of the stable manifold at the grid point, specifically, $3$ for knot, $2$ for filament, $1$ for sheet, and $0$ for void. 

Such a theoretical classification is based on different dynamical nature of large-scale structures. However, it already leads to some marked conflicts when comparing against simulations and the observed Universe even through simple visual inspections. For example, too few voids (only $\sim 17$ percent in the sense of volume occupation) were classified as shown in figure 1 of \citet{Hahn_et_al_2007}. \citet{Romero_et_al_2009} suggested that this was due to the adoption of a zero threshold, in which case the direction with even infinitesimally positive eigenvalue would be regarded as already ``collapse'', causing irrational classifications.

As investigated by \citet{Romero_et_al_2009}, $\lambda_{\rm th}$ can be theoretically approximated by equating the free-fall time $t_{\rm ff}$ of the grid to the Hubble time $t_{\rm H}$, with $t_{\rm ff} < t_{\rm H}$ ($t_{\rm ff} > t_{\rm H}$) implying collapse (expansion) of the region. Here we take the theoretical approach to obtain a rough estimate for the threshold value. Using eigenvalues of the deformation tensor and approximating the trace of the deformation tensor with $\lambda_{\rm th}$, Eq.\,(\ref{Poisson}) can be re-written as:
\begin{equation}
\label{Poisson_Eigen}
\nabla^2 \phi_{\rm m} = 4\pi G\rho_{\rm m} = \lambda_{1}+\lambda_{2}+\lambda_{3} \approx 3\lambda_{\rm th}.
\end{equation}
The free-fall time $t_{\rm ff}$ can be expressed using the local density $\rho_{\rm m}$ of the grid through:
\begin{equation}
\label{Free-fall}
t_{\rm ff} = \sqrt{\frac{3\pi}{32G\rho_{\rm m}}}.
\end{equation}
Recalling the Hubble time:
\begin{equation}
\label{Hubble}
t_{\rm H} = \frac{1}{H_0}\int_{0}^{+\infty}\frac{dz}{(1+z)E(z)},
\end{equation}
where $E(z)=\sqrt{\Omega_{\rm m}(1+z)^3+(1-\Omega_{\rm m})}$ in a flat universe dominated by matter and dark energy, and $z$ is the redshift. By setting the free-fall time $t_{\rm ff}$ equal to Hubble time $t_{\rm H}$ and through Eqs.\,(\ref{Poisson_Eigen}), (\ref{Free-fall}) and (\ref{Hubble}), we then obtain:
\begin{equation}
\label{Threshold}
\lambda_{\rm th}=\frac{\pi^2 H_0^2}{8}\left (\int_{0}^{+\infty}\frac{dz}{(1+z)E(z)}\right )^{-2}.
\end{equation}

\begin{figure}
\centering
\includegraphics[width=1\columnwidth]{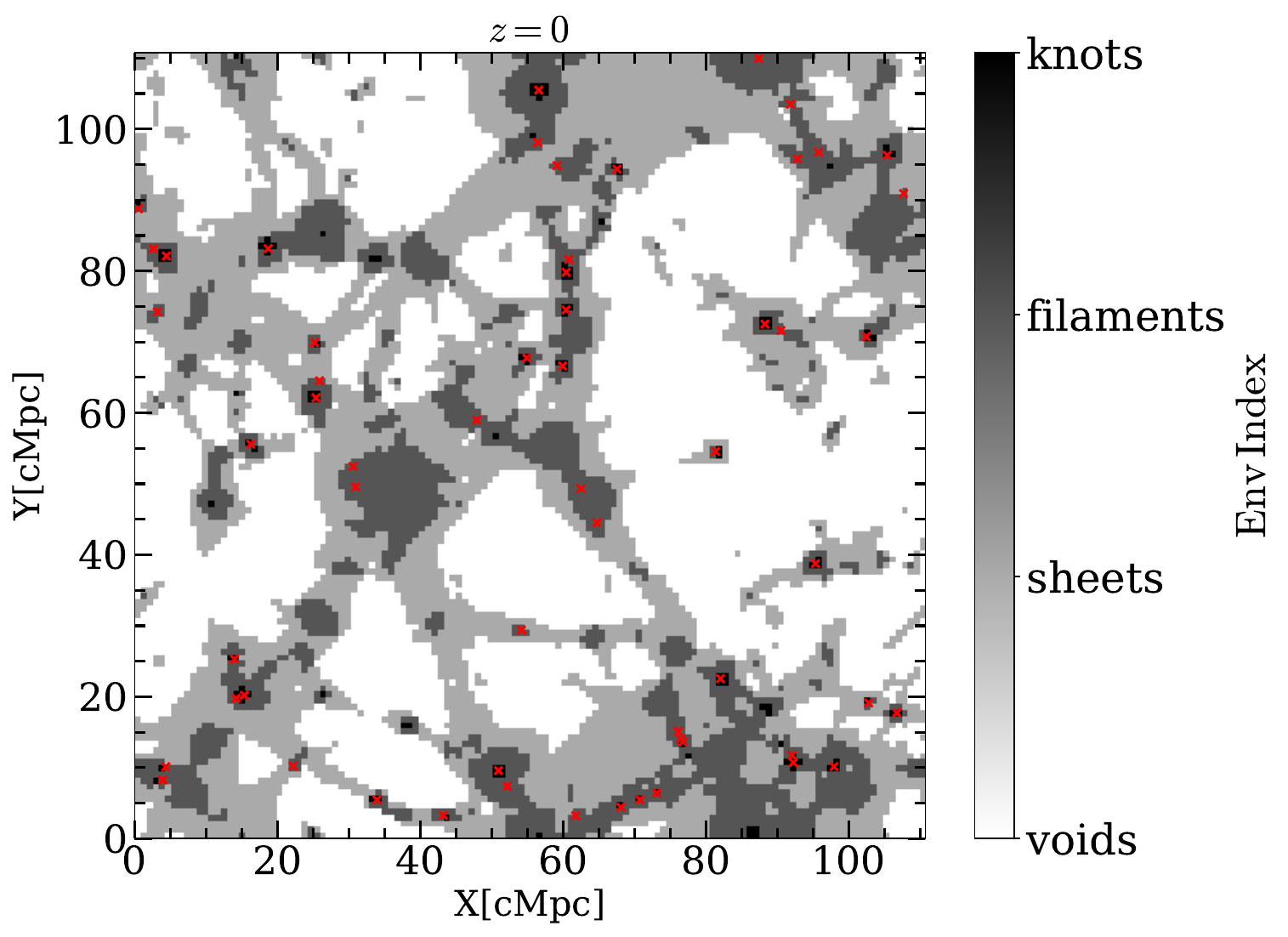}

\caption{An example slice of the large-scale structure, color-coded by the classified environment type, i.e. voids, sheets, filaments, and knots. The red crosses represent the central galaxies with halo mass larger than $10^{11.5}\mathrm{M_{\odot}}$, indicating locations of most massive galaxies in a cluster environment.}
\label{fig:Env_show}
\end{figure}

In practice through percolation analysis, \citet{Romero_et_al_2009} pinned down some effective threshold $\lambda_{\rm th}$, which corresponds to $t_{\rm ff}\approx 2.4 t_{\rm H}$, resulting in reasonable volume fractions for different types of cosmic web structures. In this work, we take one-tenth of the value given by Eq.\,(\ref{Threshold}), which yields $t_{\rm ff}\approx 3 t_{\rm H}$. The volume fractions and one example slice of the four categories are shown in Table\,\ref{table:volume_frac} and Fig.\,\ref{fig:Env_show} respectively. This result is consistent with \citet{Martizzi_et_al_2019}, who explored the fractions of different gas phases in four kinds of cosmic structures for the TNG100 simulation. We note that each central galaxy in the simulation box is then assigned with the corresponding grid type parameters as its large-scale environment type if its centre is located in that grid (i.e. via the {\it Near Grid Point} algorithm). 

\begin{figure*}
\centering
\includegraphics[width=0.6723\columnwidth]{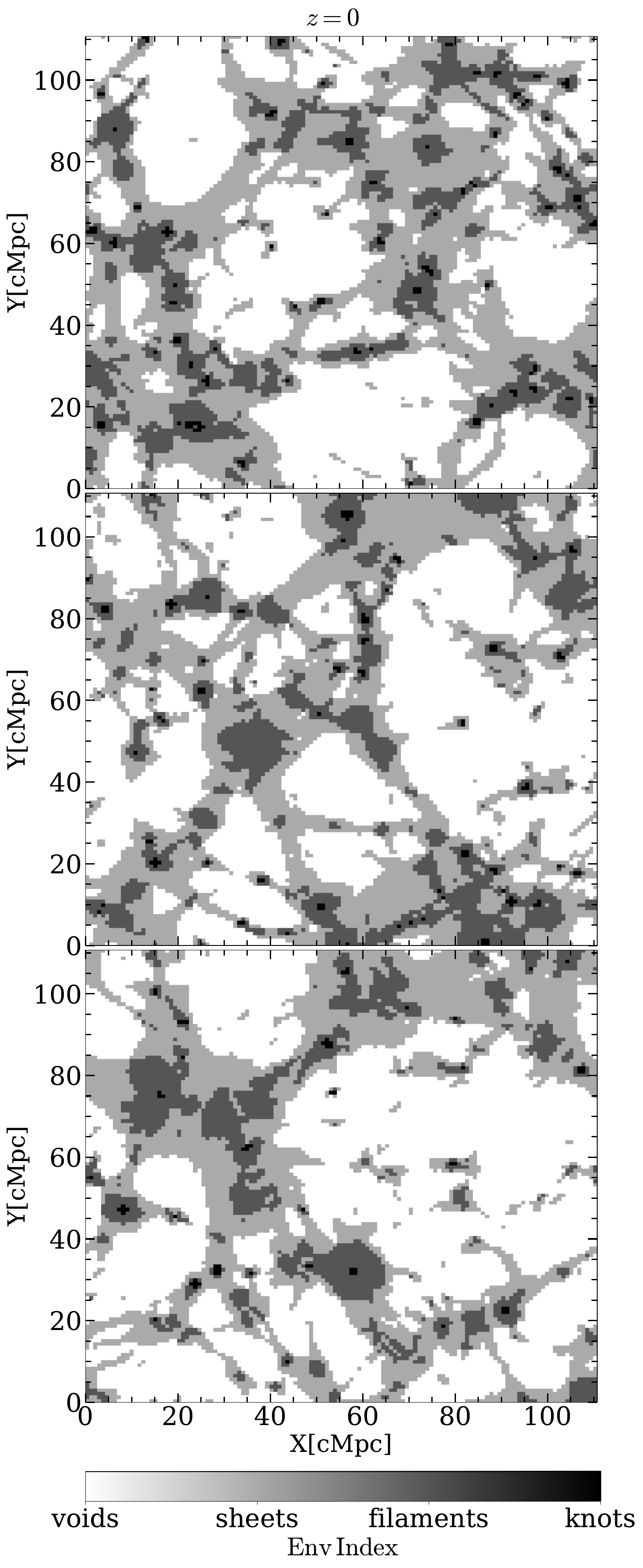}
\includegraphics[width=0.66\columnwidth]{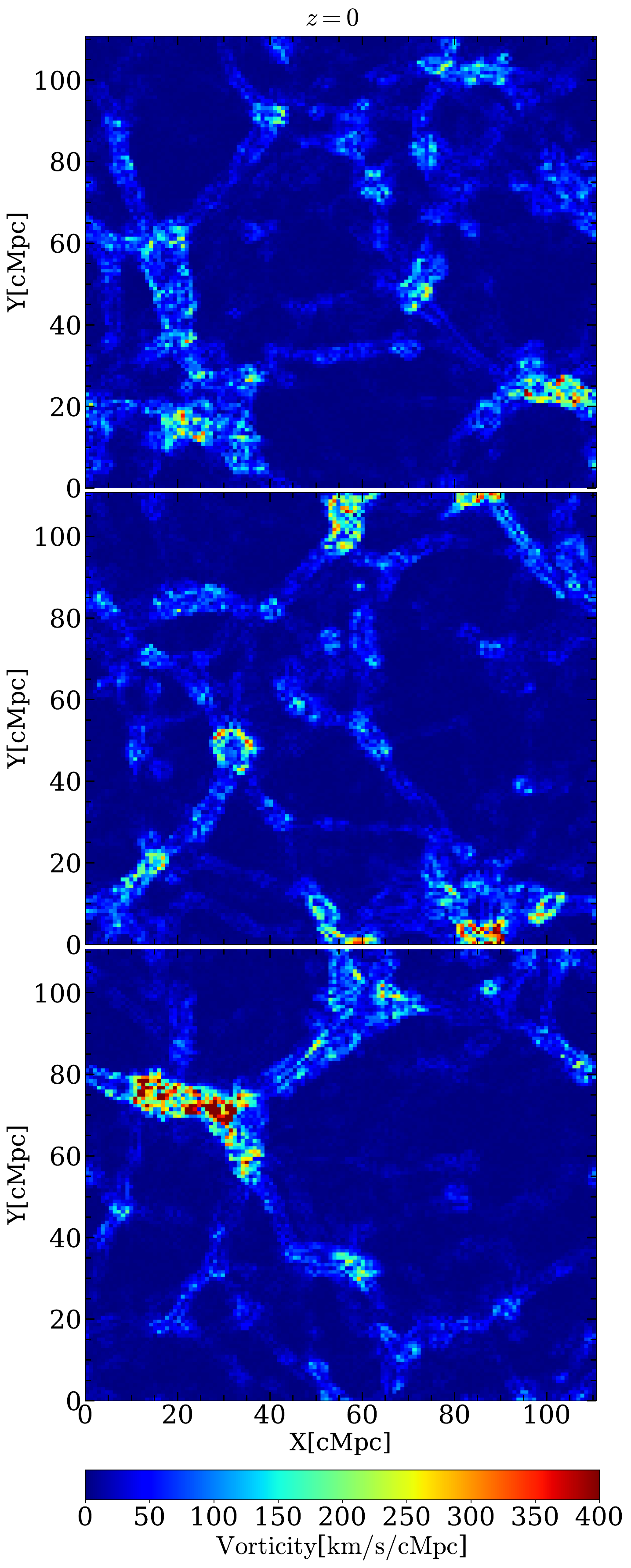}
\includegraphics[width=0.66\columnwidth]{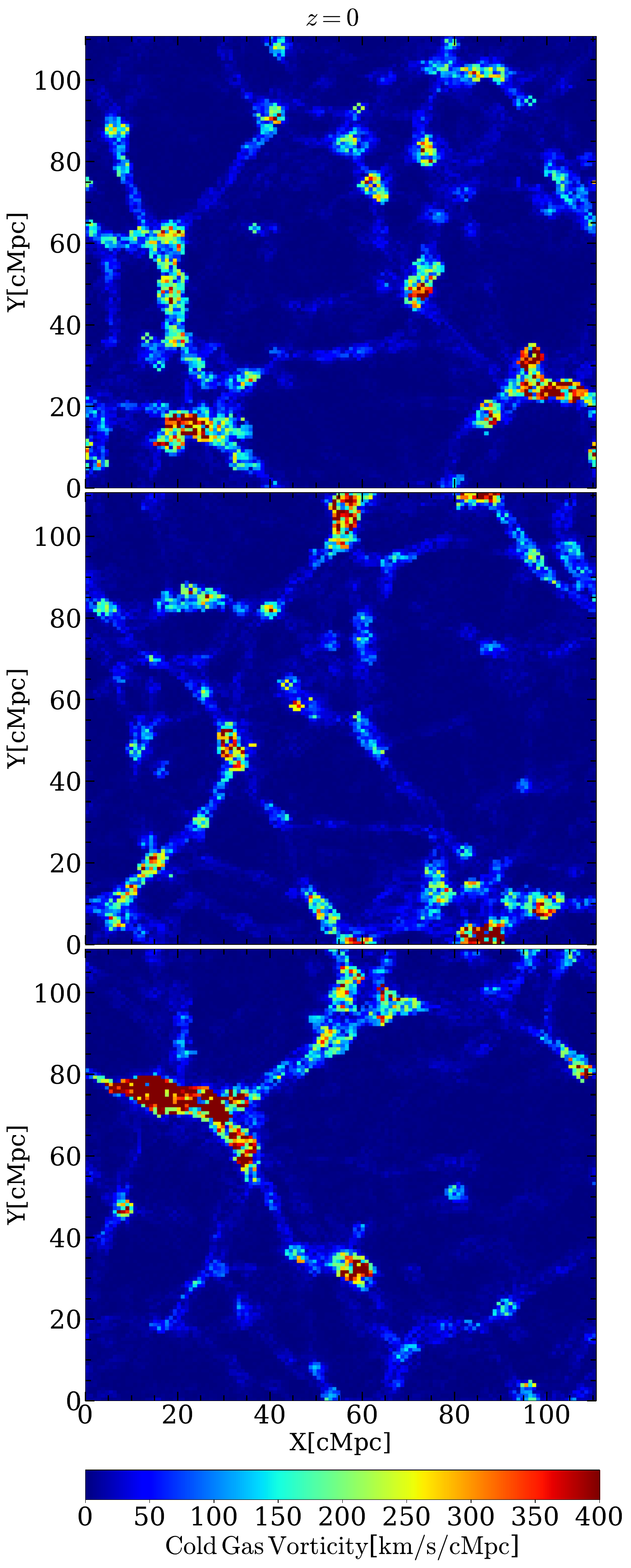}
\caption{From top to bottom are three example slices of the gridded simulation box, each with a thickness of one mesh-cell scale, i.e. $\sim 865$ kpc, and from left to right, color-coded by the corresponding environment type, total matter vorticity, and cold gas vorticity, respectively.} 
\label{fig:Vorticity_Example}
\end{figure*}

In Fig.\,\ref{fig:Vorticity_Example} we present three example slices in the simulation box, each with a thickness of one mesh-cell scale, i.e. $\sim 865$\,kpc. From left to right, the fields are color-coded by the corresponding environment type, total matter vorticity, and cold gas vorticity, respectively. In Section \ref{sec:general_sample}, we will present how the cold-gas vorticity depends on the environmental type, as well as the relation between the cold-gas vorticity and central sSFR in different large-scale environments.

\begin{table}
\caption{The volume fraction of four categories at $z\sim 0$.} 
\setlength{\tabcolsep}{6mm}
\begin{tabular}{cccc}
\hline
\hline
$\rm Void$ & $\rm Sheet$ & $\rm Filament$ & $\rm Knot$\\
\hline
53.75\% & 33.42\% & 12.29\% & 0.54\%\\
\hline
\end{tabular}
\vspace{2mm}
\label{table:volume_frac}
\end{table}

\section{Cold-Gas Vorticity Environment among Typical Galaxy Samples} 
\label{sec:old}

\begin{figure*}
\centering
\includegraphics[width=2\columnwidth]{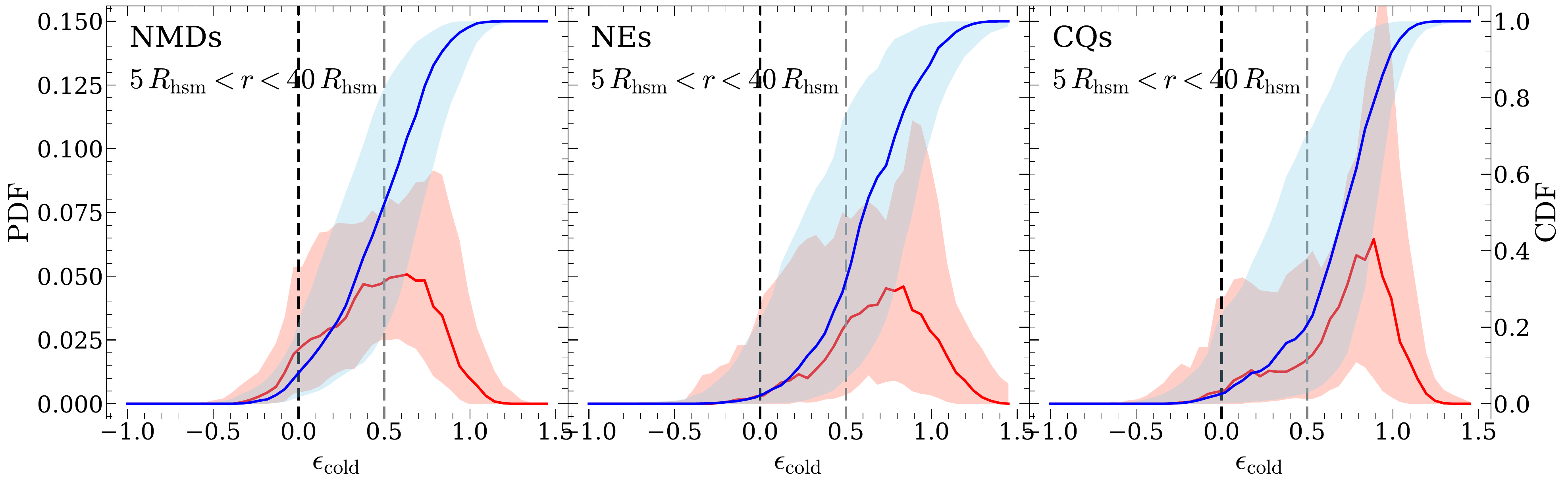}
\includegraphics[width=2\columnwidth]{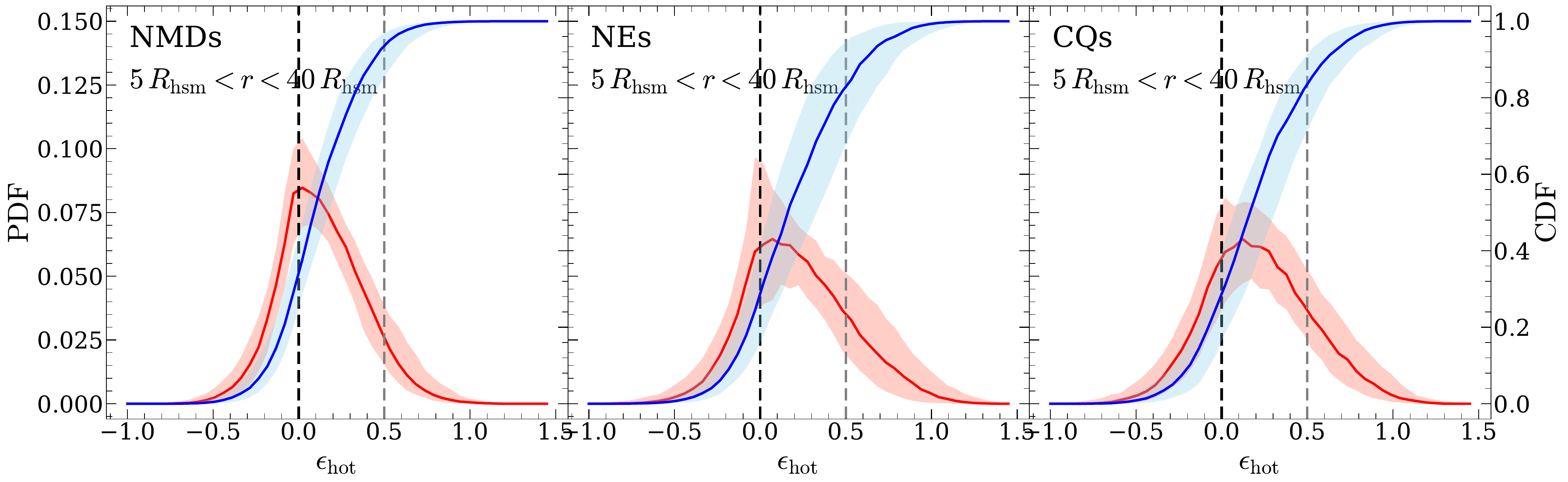}
\caption{Circularity distributions of the CGM gas among three types of galaxy samples at $z\sim 0$ within $5R_{\rm hsm}<r<40R_{\rm hsm}$. {\bf NMD} stands for normal star-forming disc galaxies, {\bf CQ} stands for dynamically-cold but quenched early-type galaxies, and {\bf NE} represents for dynamically-hot but quenched early-type galaxies (i.e., normal ellipticals). The upper row is for the cold CGM and the lower row shows the hot CGM. Red lines represent the probability distribution functions (PDFs) in gas mass and the blue lines are the corresponding cumulative distribution functions (CDFs), with shaded regions indicating the $1\sigma$ ranges. Black and grey dashed lines mark the circularity of 0 and 0.5 respectively. The larger the circularity, the closer the gas cell to circular orbit.}
\label{fig:Circularity}
\end{figure*}

\begin{figure*}
\centering
\includegraphics[width=1\columnwidth]{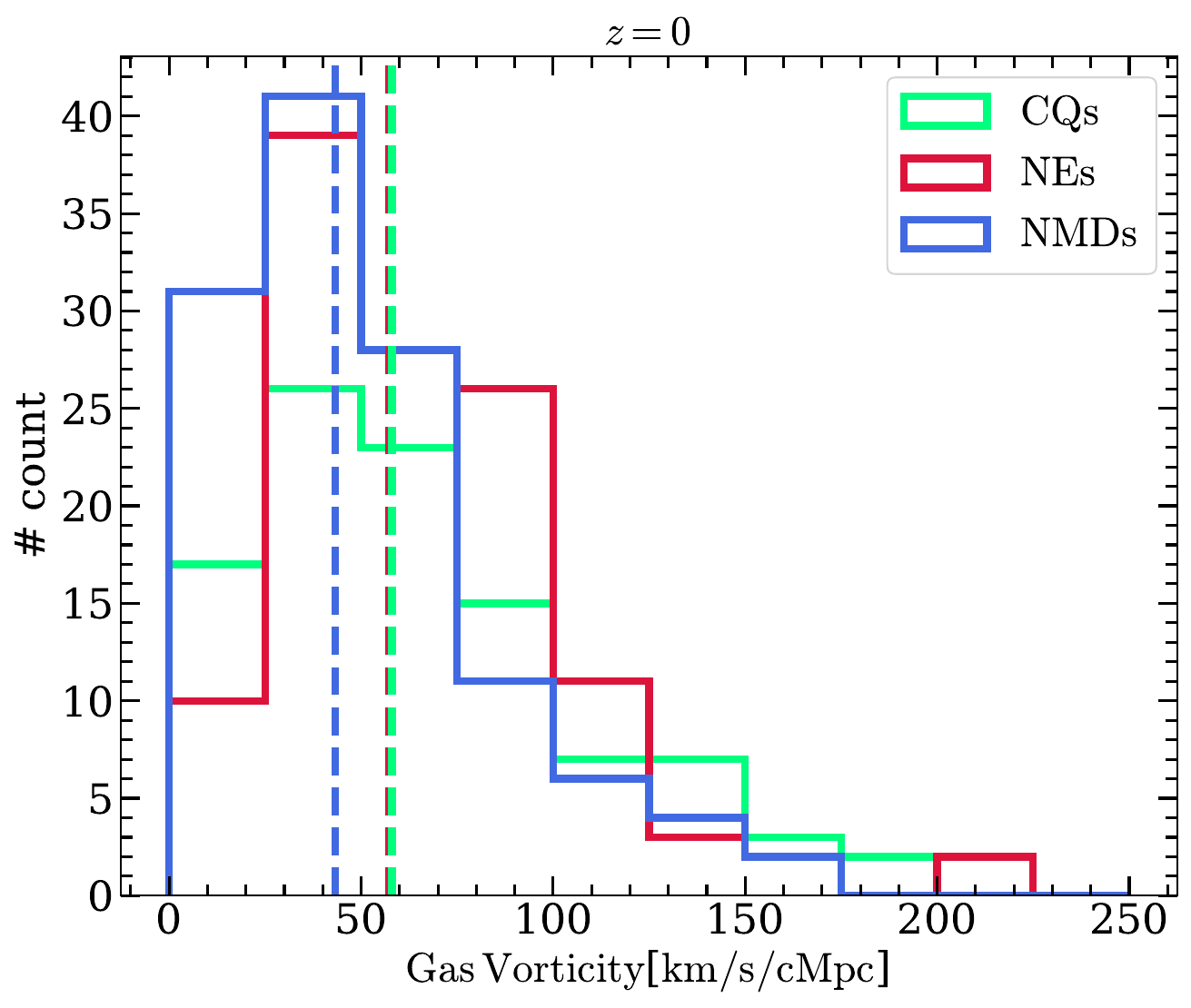}
\includegraphics[width=1\columnwidth]{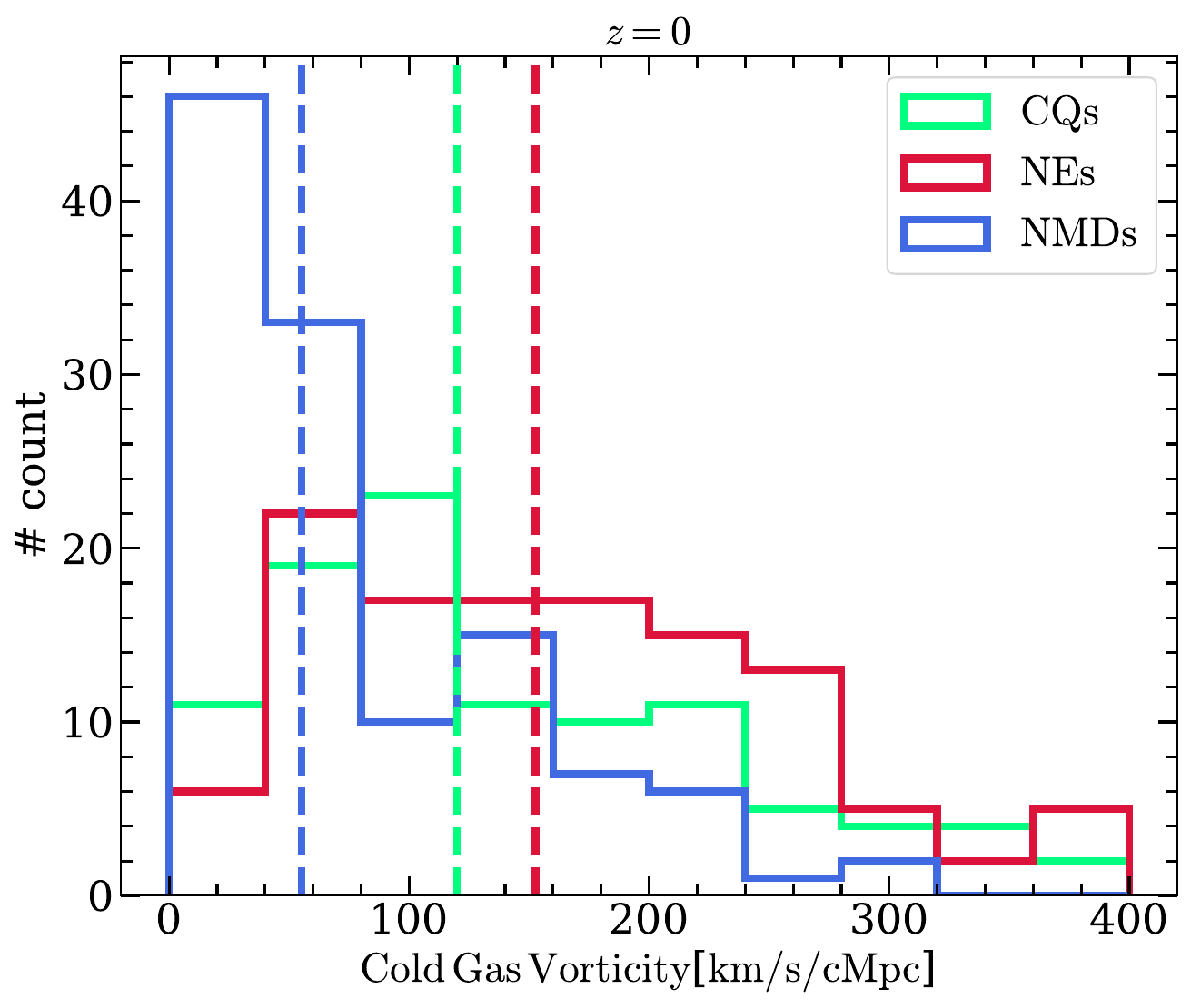}
\caption{Gas vorticity (left) and cold gas ($T_{\rm gas} < 2\times 10^4\,\mathrm{K}$) vorticity (right) distributions among three types of samples at $z\sim 0$. {\bf NMD} (in blue) stands for normal star-forming disc galaxies, {\bf CQ} (in green) stands for dynamically-cold but quenched early-type galaxies, and {\bf NE} (in red) represents for dynamically-hot but quenched early-type galaxies (i.e., normal ellipticals). }
\label{fig:Vorticity}
\end{figure*}

This study is an extension of the previous works \citep{Wang_et_al_2022, Lu_et_al_2022}, where we studied the angular momentum modulation of the environment to the CGM kinematics and star formation activities using about three hundreds typical galaxies from the TNG100 simulation. 
As we have stated in the introduction, it is the cold gas accretion that is directly relevant to the central star forming activities, we therefore first demonstrate the different motion of the cold gas with respect to the hot component, in the case of the three hundreds galaxies from our previous studies. 


These galaxies (within a similar stellar mass range of $10.3 < \log\,M_{\ast}/\mathrm{M_{\odot}} < 11.2$ ) from our previous studies fall into three categories, i.e., normal {\bf star-forming} disc galaxies, dynamically-{\bf cold} but {\bf quenched} early-type galaxies and  dynamically-{\bf hot} but {\bf quenched} early-type galaxies. 
Following figure 6 in \citet{Lu_et_al_2022}, here we plot the circularity (for definition, see Equation 3 therein) distributions of the CGM gas in the three types of galaxies in Fig.\,\ref{fig:Circularity}, but splitting the gas into cold and hot components. The thresholds for the cold and hot gas are $T_{\rm gas} = 10^4\,\mathrm{K}-2 \times 10^4\,\mathrm{K}$ and $T_{\rm gas} > 10^5\,\mathrm{K}$ respectively, same as in \citet{Wang_et_al_2022}. Such a gas circularity defined for each gas cell is essentially the ratio between the specific angular momentum (of this gas cell) projected along the total CGM spin direction (calculated within $5R_{\rm hsm}<r<40R_{\rm hsm}$) and the maximum angular momentum of a steady Keplerian orbit at that distance, roughly indicating the strength of rotation. 

It can be seen that in all three types of galaxies the cold CGM gases always favor higher angular momenta with peaks at $\epsilon > 0.5$ in the gas circularity distributions, in obvious contrast to the hot CGM gases, for which the $\epsilon$ distributions peak around zero. The mass fractions of the cold CGM gases that have $\epsilon > 0.5$ are $\sim 48$ per cent, $68$ per cent and $79$ per cent for the normal {\bf star-forming} disc galaxies, dynamically-{\bf cold} but {\bf quenched} early-type galaxies, and dynamically-{\bf hot} but {\bf quenched} early-type galaxies, respectively. The fact that the quenched populations (latter two) generally have larger fractions of their {\it cold} gas components possessing higher angular momenta in comparison to star-forming galaxies, suggests that the angular/tangential motion possibly prevent cold gas from efficient infalling to smaller radii to fuel central star formation. 
  

The vorticity $\boldsymbol{\omega}$ defined as the curl of a velocity field may to some extent indicate the rotational strength of the local matter flow. 
In Fig.\,\ref{fig:Vorticity}, we show the vorticity distributions of the ($\sim$ 1\,Mpc)$^3$ cell environment around the three types of galaxy samples, with the left panel presenting the vorticity of total gas and the right panel the cold gas. It is clear that 
differences in the cold-gas vorticities among the three galaxy populations are much more significant than those in the total gas vorticity. In particular, the quenched galaxy samples exhibit much larger cold-gas vorticities than the star-forming samples. 
Again the star-forming samples live in the environment with lower cold-gas vorticities, conducive to more efficient cold-gas inflow to fuel central star formation, while the opposite is true for the quenched galaxy populations.


\section{Cold Gas Vorticity Environment for a General Sample of Central Galaxies} 
\label{sec:general_sample}

In this section, we mainly focus on the cold-gas vorticity properties, because star formation activities are directly related to the cold gas accretion (e.g., \citealt{Keres2005, Dekel2006, Dekel2009Nature}). To do so, we have enlarged the galaxy samples to all central galaxies that have stellar masses with $10.0<\log\,M_{\ast}/\mathrm{M_{\odot}}<11.5$,
and explored the influence of the large-scale vorticity on galaxy star-formation properties. The lower mass cut is introduced to guarantee sufficient resolution, while the upper mass cut is to, at the sample selection stage, already exclude galaxies whose quenching statuses are highly susceptible to the influence of AGN feedback. \citet{Pillepich_et_al_2018b} studied TNG100 galaxies in group and cluster environments and demonstrated that the most massive galaxies with stellar masses $M_{\ast}>3\times 10^{11}\, \mathrm{M_{\odot}}$ are located in cluster halos of total masses beyond $10^{14}\,\mathrm{M_{\odot}}$. The evolution history of these most massive galaxies differ from their less massive counterparts in two significant ways. Firstly, they build up a significant fraction as high as 80\% of their stellar population through cosmic merger and accretion of lower mass galaxies; even within the innermost 10 kpc, the fraction of ex-situ stars can be as high as 60\% (see also \citealt{Rodriguez-Gomez2017}). This is a consequence of a markedly enhanced galaxy merger rate at this mass scale (\citealt{Hopkins2010Merger, Rodriguez-Gomez2015}). 

In addition, these most massive galaxies have suppressed star formation rates largely due to the action of supermassive black hole feedback which generates a large fraction of outflow gas through direct ejection of galaxy central gas (\citealt{Weinberger2017, Nelson_et_al_2019b}). The outflow gas has a temperature peak at a few times $10^4$\,K, and reaches an outward-moving velocity beyond 1000 km/s (with a total flux of $10\, \mathrm{M_{\odot}/yr}$). As a consequence, these most massive galaxies exhibit a clear dented feature in their SFR surface density profiles in the central regions (\citealt{Nelson_et_al_2019b}). Clearly, the central star formation in this case is strongly suppressed by their supermassive black hole activities, rather than regulated by the ambient cold-gas motion that affects central gas acquisition. This is the reason that we set an upper limit on the galaxy mass. In fact, AGN feedback already effectively takes action in galaxies with stellar masses beyond $10^{11}\, \mathrm{M_{\odot}}$, corresponding to halo masses beyond $10^{13}\, \mathrm{M_{\odot}}$ (see \citealt{Pillepich_et_al_2018b, Nelson_et_al_2019b}). As we will see in this section, the effect of AGN feedback on the cold gas vorticity indeed shows up for galaxies living in halos more massive than $10^{13}\, \mathrm{M_{\odot}}$, we present detailed discussion in this section.

\subsection{Higher cold-gas vorticity corresponds to lower sSFR}
\label{ssfr_vorticity}

\begin{figure}
\centering
\includegraphics[width=1\columnwidth]{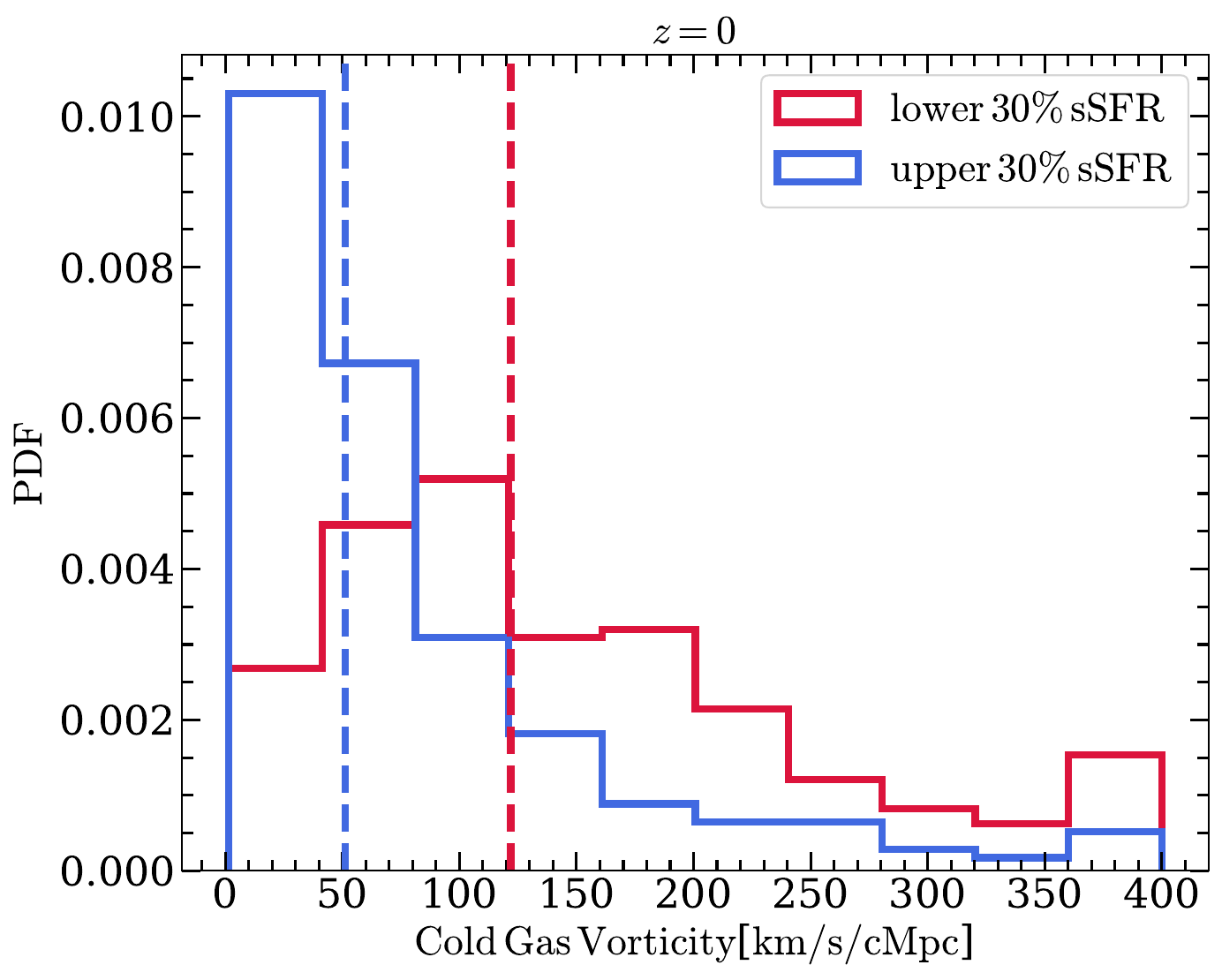}
\includegraphics[width=1\columnwidth]{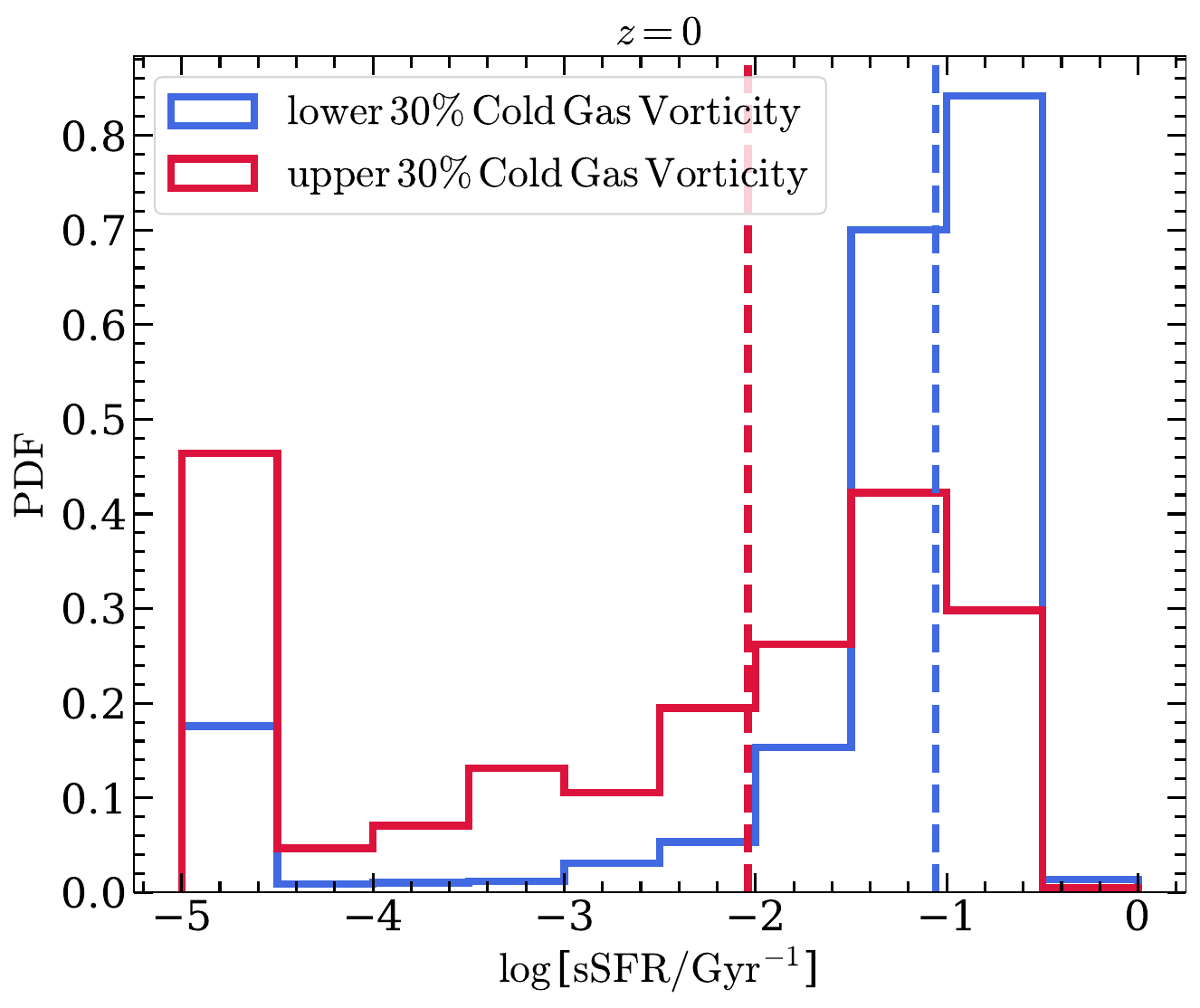}
\caption{Upper: The cold gas vorticity distributions of galaxies with top (blue) and bottom (red) 30 per cent sSFR. Lower: The sSFR distributions of galaxies with top (red) and bottom (blue) cold gas vorticity. All sample galaxies are centrals with stellar mass in $10^{10} {\rm M_{\odot}} - 10^{11.5} {\rm M_{\odot}}$ at $z\sim 0$.}
\label{fig:Cold_Vorticity}
\end{figure}

To examine the angular momentum modulation, all central galaxies at $z=0$ and with stellar masses ranging from $10^{10} {\rm M_{\odot}}$ to $10^{11.5} {\rm M_{\odot}}$ are considered. We particularly focus on the comparison between two sub-groups of these galaxies, one among the top 30 per cent and the other among the bottom 30 per cent of a given property. We then examine the distributions of other properties and make comparisons among the two sub-groups. In Fig.\,\ref{fig:Cold_Vorticity}, the upper left panel shows the cold gas vorticity distributions of galaxies that have the highest (in blue) and the lowest (in red) 30 per cent sSFR (i.e. the most and least star-forming samples). As is shown, statistically more (least) actively star-forming galaxies live in an environment with lower (higher) cold-gas vorticities. When we reverse these two properties and compare the sSFR distributions of the two galaxy samples that live in environments which possess the upper (in red) and lower (in blue) 30 per cent of cold-gas vorticities, as is presented in the lower panel, we also find that the environment that has lower cold-gas vorticities tend to host more actively  star-forming galaxies in comparison to its higher vorticity counterpart. 

Fig.\,\ref{fig:Cold_Vorticity} is necessary to support our angular momentum scenario but not sufficient, because we have not ruled out other factors that might degenerate with the vorticity. For example, we know that both the halo spin and the level of quenching increase with the dark matter halo mass. If the vorticity is also positively correlated with the halo mass, then the suppress of the star formation activity in the high vorticity case could simply because galaxies tend to be more quenched towards higher masses. On the one hand, kinematically, whether the material can flow into the galaxy center or not is partially determined by the strength of the gravitational pull (at given angular momentum). On the other hand, halos with different masses will heat up the CGM to different degrees, resulting in different cooling efficiencies and thus star formation activities. For this reason, we shall also take into account the dependence on halo mass.

\begin{figure}
\centering
\includegraphics[width=1\columnwidth]{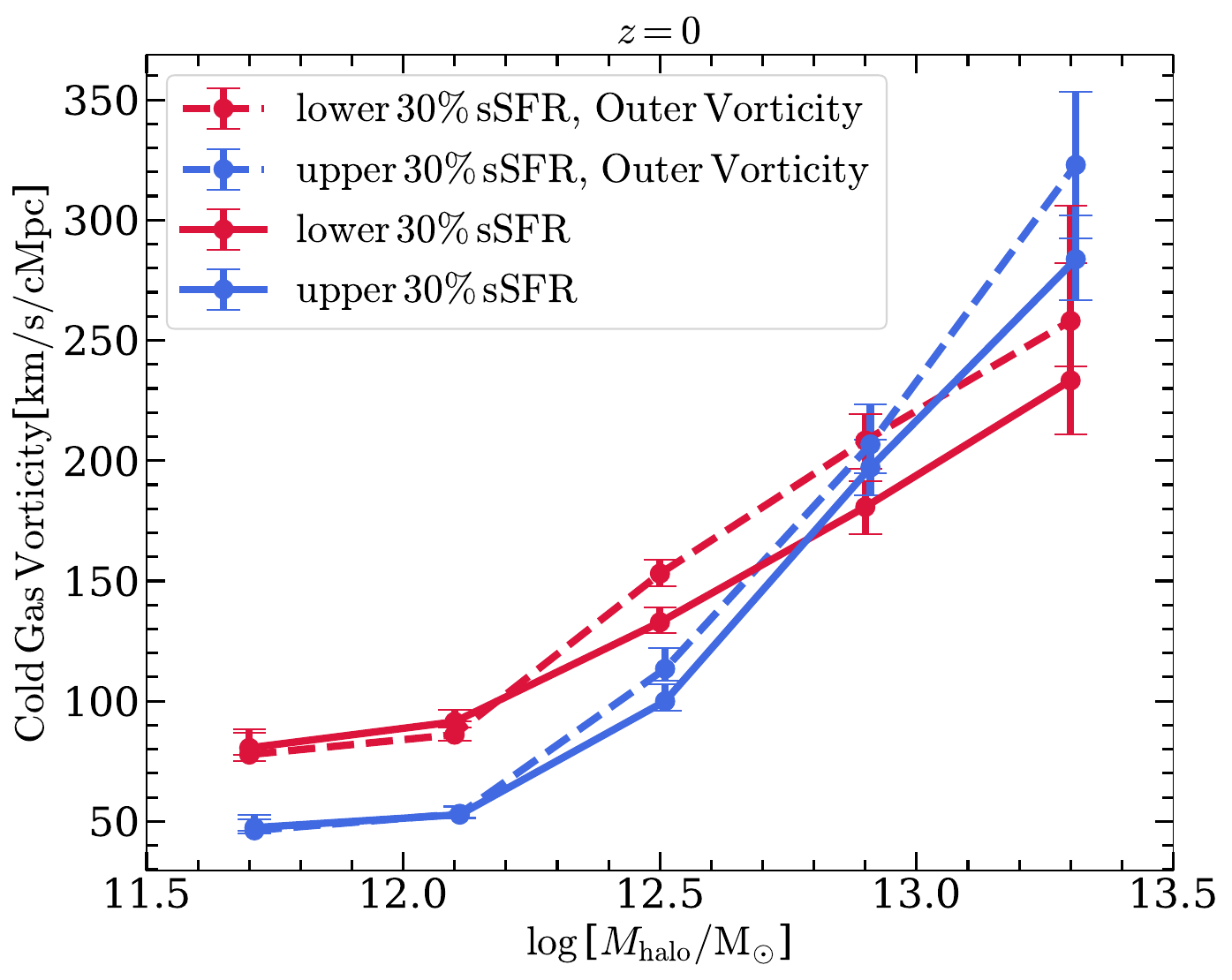}
\includegraphics[width=1\columnwidth]{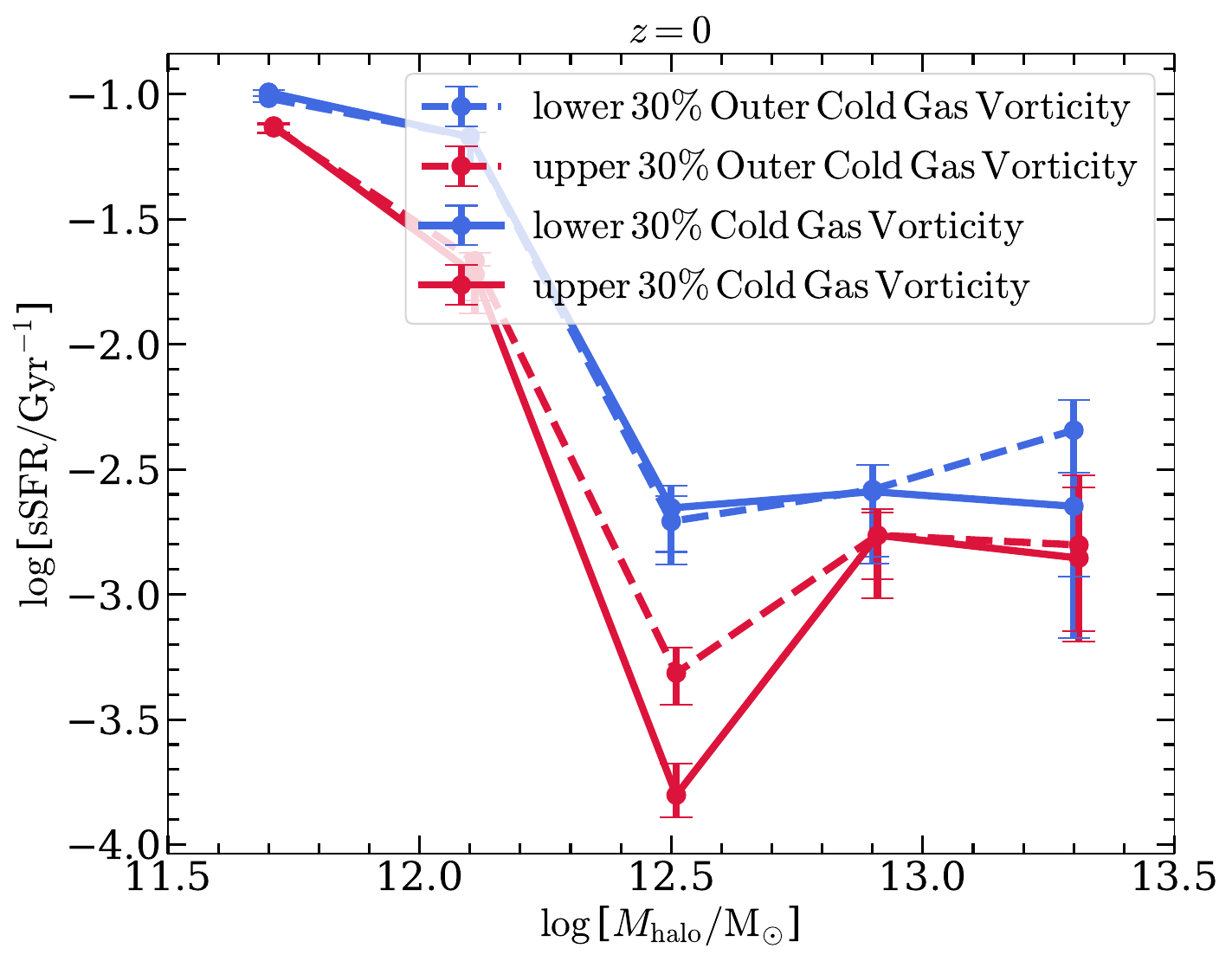}
\caption{Upper: The cold gas vorticity distributions of galaxies with top (blue) and bottom (red) 30 per cent sSFR \textit{in each halo mass bin}. Lower: The sSFR distributions of galaxies with top (red) and bottom (blue) 30 per cent cold gas vorticity \textit{in each halo mass bin}. Data points show the median values and error bars indicate the standard error of the mean in each bin. Solid and dashed lines represent the results for vorticity of all cold gas and cold gas excluding any halo's central part, i.e. 1/3 $r_{200}$.}
\label{fig:ssfr_Vorticity_Binned}
\end{figure}

Fig.\,\ref{fig:ssfr_Vorticity_Binned} shows the cold gas vorticity (upper panel) and the sSFR (lower panel) as a function of halo mass $M_{\rm halo}$, which refers to the spherical overdensity mass $M_{200}$ which is the total mass within $r_{200}$ of the dark matter halo of the corresponding galaxy, as identified by {\sc subfind}. We split galaxies into five mass bins, i.e., $\log\,M_{\rm halo}/\mathrm{M_{\odot}}\,\in\,[11.5,11.9],\,[11.9,12.3],\,[12.3,12.7],\,[12.7,13.1],\,[13.1,13.5]$. Within each mass bin, we split the galaxies again into those with the top or bottom 30 per cent of a given property, i.e., either the sSFR or the environmental cold-gas vorticity. The points connected by lines represent the median values of galaxies in the two sub-groups in a given halo mass bin. The error bars label the standard error of the mean in each bin.  
From the upper panel, the trend that the higher (lower) the galaxy sSFR the lower (larger) the cold gas vorticity is clearly seen across the first three lower-mass bins at below $\log [M_{\rm halo}/{\rm M_\odot}] \sim 13$. 
At the higher mass end above $\log [M_{\rm halo}/{\rm M_\odot}] \sim 13$, the results go opposite. We note the reader that despite of having already adopted a stellar mass cut at the high mass end to pre-exclude galaxies that are most affected by AGN feedback, the AGN feedback already takes action in these more massive galaxies, where the central gas (including a significant amount of cold gas) can be ejected at a velocity of 100-1000 km/s out to distances of a few hundred kpc (see \citealt{Weinberger2017, Nelson_et_al_2019b}). This enhances the radial motion of gas kinematics at larger distances, leading to a lower vorticity environment for these most massive and passive galaxies. 

In the lower panel of Fig.\,\ref{fig:ssfr_Vorticity_Binned}, we plot the sSFR as a function of halo mass for galaxies that live in environments with upper and lower 30 per cent of the cold-gas vorticity within each mass bin. Again we see that higher (lower) cold gas vorticity environment tends to host more quenched (star-forming) galaxies. Specifically, at a fixed halo mass scale of $10^{12}-10^{13}\,\mathrm{M_{\odot}}$, the median sSFR of galaxies living in an environment with the top 30\% cold-gas vorticity is $\sim 0.5$ dex below that of galaxies living in an environment with the bottom 30\% cold-gas vorticity.


It is worth noting that the negative trend between the cold gas vorticity and the sSFR exhibited here is consistent with the results from our previous studies (\citealt{Wang_et_al_2022, Lu_et_al_2022}), where a negative correlation was found between the CGM spin and the sSFR. However, such a correlation could be erased or altered, because both the CGM gas motion in the inner parts of halos and the star formation activities can be strongly modulated by various internal feedback effects (as already present here in the highest mass bins). In particular, the trend between the two that we discovered in our previous studies using the TNG100 simulation could be different using other cosmological simulations which take different approaches to implement feedback processes. In \citet{Liu_et_al_2024}, the authors compared such correlation derived from the SIMBA simulation (\citealt{Dave2019SIMBA}) with that from TNG100 and found the opposite trend, e.g., quenched galaxies have lower CGM spin. In order to eliminate effects from any possible internal modulation, we have also excluded cold gases within the central $r_{200}/3$ of each halo (shown in dashed lines) and compared the results with those when including all cold gases (shown in solid lines). As can be seen, when {\it only} taking into account the cold gases at larger distances from halo centres, not only our findings remain present, but also the flip of the trend (in the top panel where the blue and red lines cross) moves to higher mass bins. This indicates that the conclusion on a negative modulation from the ambient cold gas vorticity to galaxy star formation activeness that we report here is not dominated by the interstellar medium or the inner circumgalactic gas, but a result from larger distances. At below $\log [M_{\rm halo}/{\rm M_\odot}] \sim 13$, such a conclusion is less susceptible to various internal feedback effects.


\subsection{Star-forming versus Quenched}
\label{starforming_quenched}
In the former sections, we illustrate the negative impact of high environment angular momentum and high cold gas vorticity to the galaxy star formation activeness. One thing that we shall  emphasize is that we do not propose this mechanism to account for the significant shutdown of the past star formation in present-day quenched galaxies. 
For these galaxies, the angular momentum mechanism only serves in the ``post-quenched'' phase, playing a role of preventing the system from  efficient cold-gas supplement to galaxy centres and thus further star formation, when the environmental angular momentum is high. In fact, if it is otherwise, i.e., the environmental gas angular momentum is not sufficiently high, then galaxies might rejuvenate, moving back to the blue cloud from the green valley (or even the red sequence), back and forth, experiencing episodic star formation history, either before a complete shutdown of star formation in present-day quenched galaxies (e.g., \citealt{Tacchella_et_al_2016,Behroozi_et_al_2019,Wang_et_al_2019}), or throughout the late-time evolution of present-day star-forming galaxies (see figure 1 of \citealt{Wang_et_al_2022}). 

\begin{figure*}
\centering
\includegraphics[width=1\columnwidth]{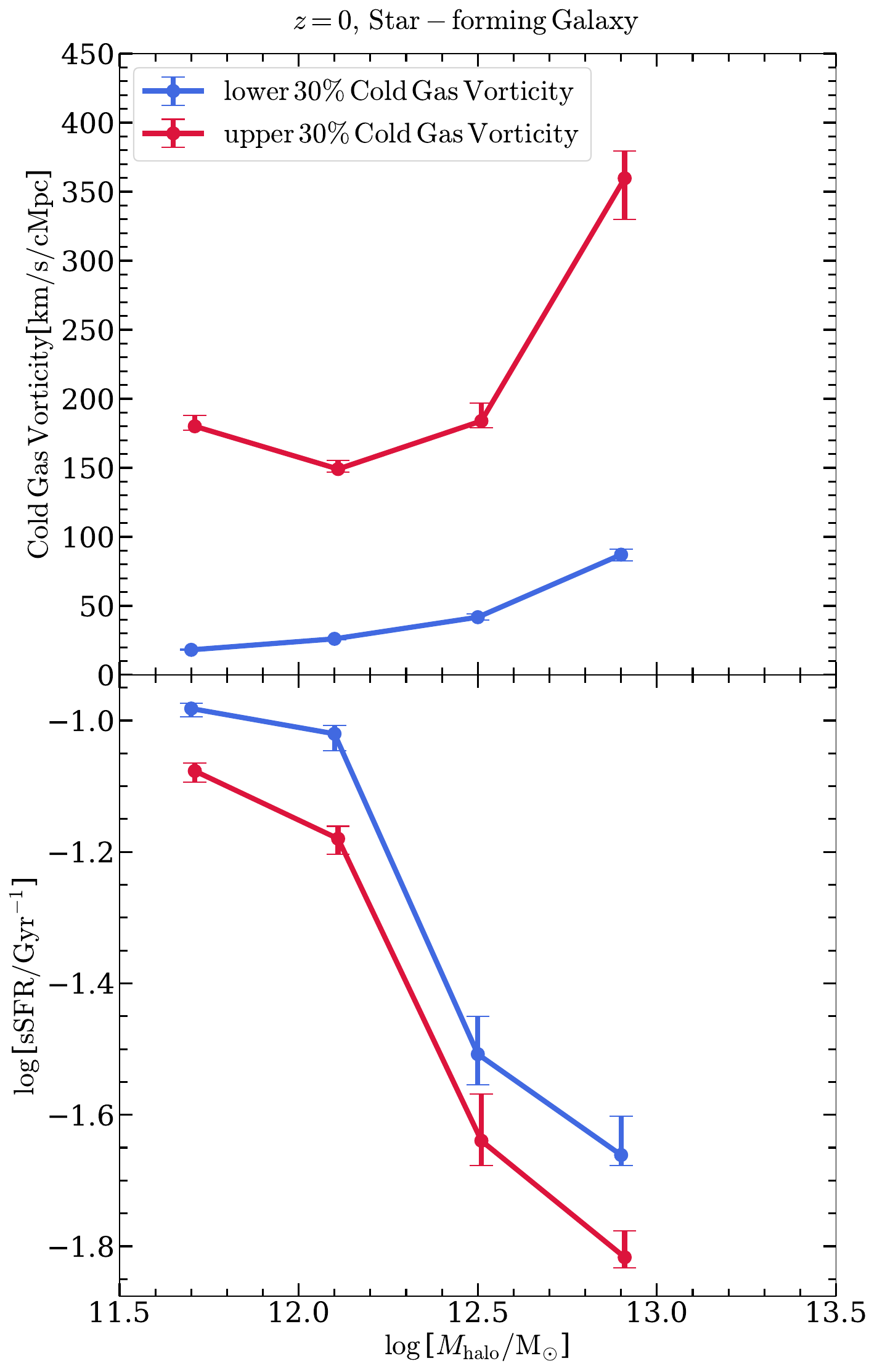}
\includegraphics[width=1\columnwidth]{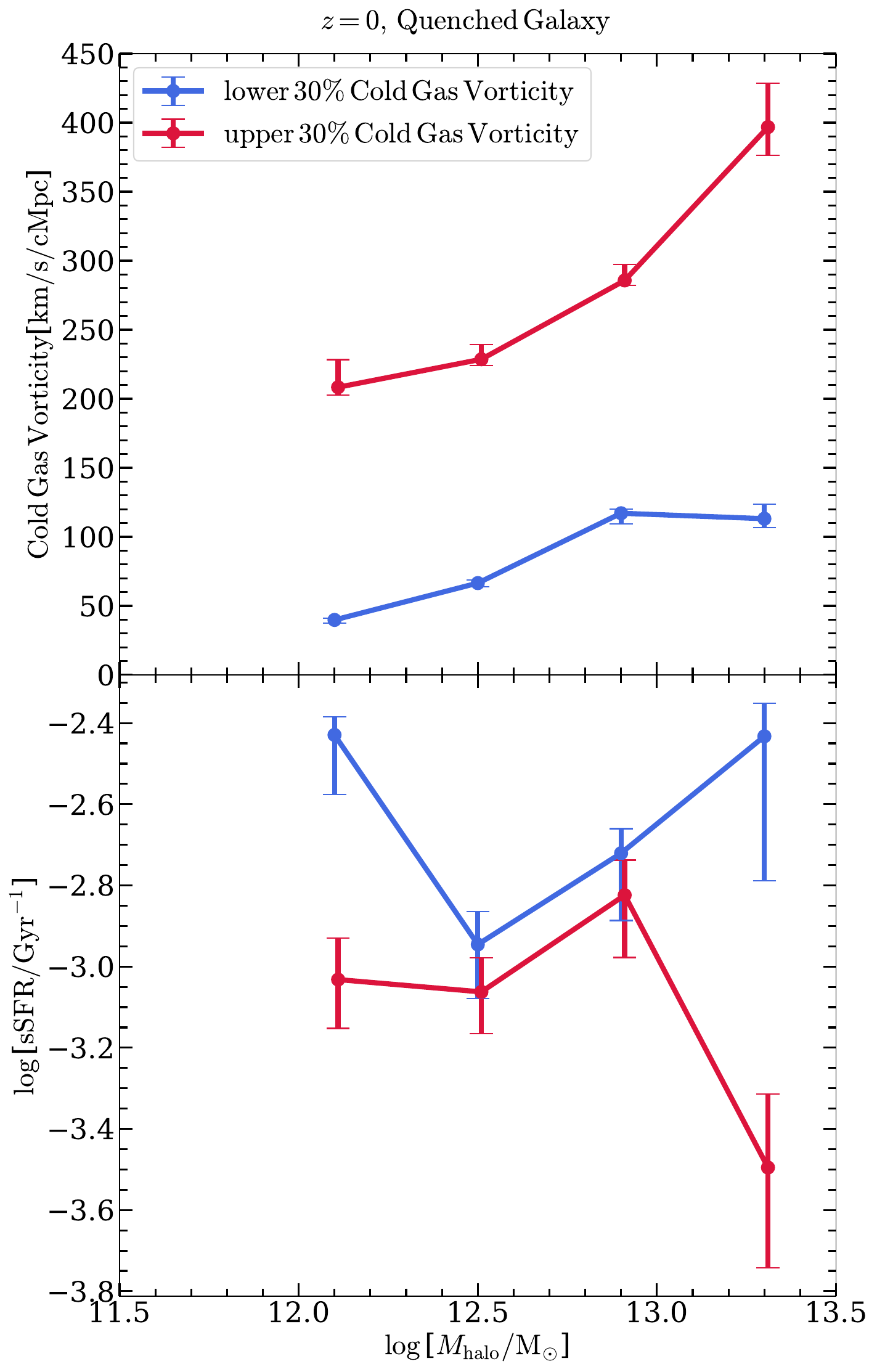}
\caption{The cold gas vorticity (top) and sSFR (bottom) of galaxy environments that possess top 30 per cent (red) and bottom 30 per cent (blue) of the cold gas vorticity {\it within each halo mass bin}. Left panels are for star-forming galaxies and right ones are quenched samples. Galaxies that are totally quenched, i.e. with $\log [\rm sSFR/Gyr^{-1}] <= -5$, are ignored here.}
 
\label{fig:ssfr_Vorticity_Binned_SF_QG}
\end{figure*}

A further clarification is that the angular momentum mechanism does not only takes effect after galaxies being quenched. The low/high angular momentum environment is always there and universally affects the gas acquisition efficiency and thus SFR for all types of galaxies, also including star-forming galaxies. To demonstrate this, we dichotomously divide galaxies into star-forming and quenched samples, as separated by $\log [{\rm sSFR}/{\rm Gyr}^{-1}] = -2.0$ (e.g., \citealt{Lu_2022_ColdQuenched}). For either galaxy population, we plot the sSFR distribution as a function of halo mass, as is shown in Fig.\,\ref{fig:ssfr_Vorticity_Binned_SF_QG}, and make comparisons between the two sub-groups of galaxies that live in the top and bottom 30 per cent cold-vorticity environments within each halo mass bin. We note that in the right panels of Fig.\,\ref{fig:ssfr_Vorticity_Binned_SF_QG} (i.e., for quenched galaxies), we have dismissed galaxies that are totally quenched (i.e. $\log [\rm sSFR/Gyr^{-1}] <= -5$) in order to reduce the contamination due to the much stronger quenched mechanisms related to baryonic feedback processes. As can be seen, for either galaxy population and within each halo mass bin, galaxies that live in environments with larger (smaller) cold-gas vorticities (as presented in the top panels) always tend to have lower (higher) sSFR (as shown in the bottom panels). It is clear that a negative modulation of angular momentum to star formation activeness has been widely observed among both star-forming and quenched galaxy populations. 



\subsection{Dependence on Larger-Scale Environment Type}
\label{dep_Env_typ}
As already introduced in Section \ref{sec:introduction}, past studies already revealed that the angular momentum environment, in particular the vortical flow associated with different large-scale environments acts on different levels. In this work, we have also investigated how the environmental cold-gas vorticity and its influence on galaxy star formation would differ among different large-scale environmental types. In particular, we make comparisons between situations in filaments and knots. As is shown by \citet{Martizzi_et_al_2019}, filaments and knots in general tend to host much higher fractions of cool and dense gas than sheets and voids (see figure 5 therein).    

\begin{figure}
\centering
\includegraphics[width=1\columnwidth]{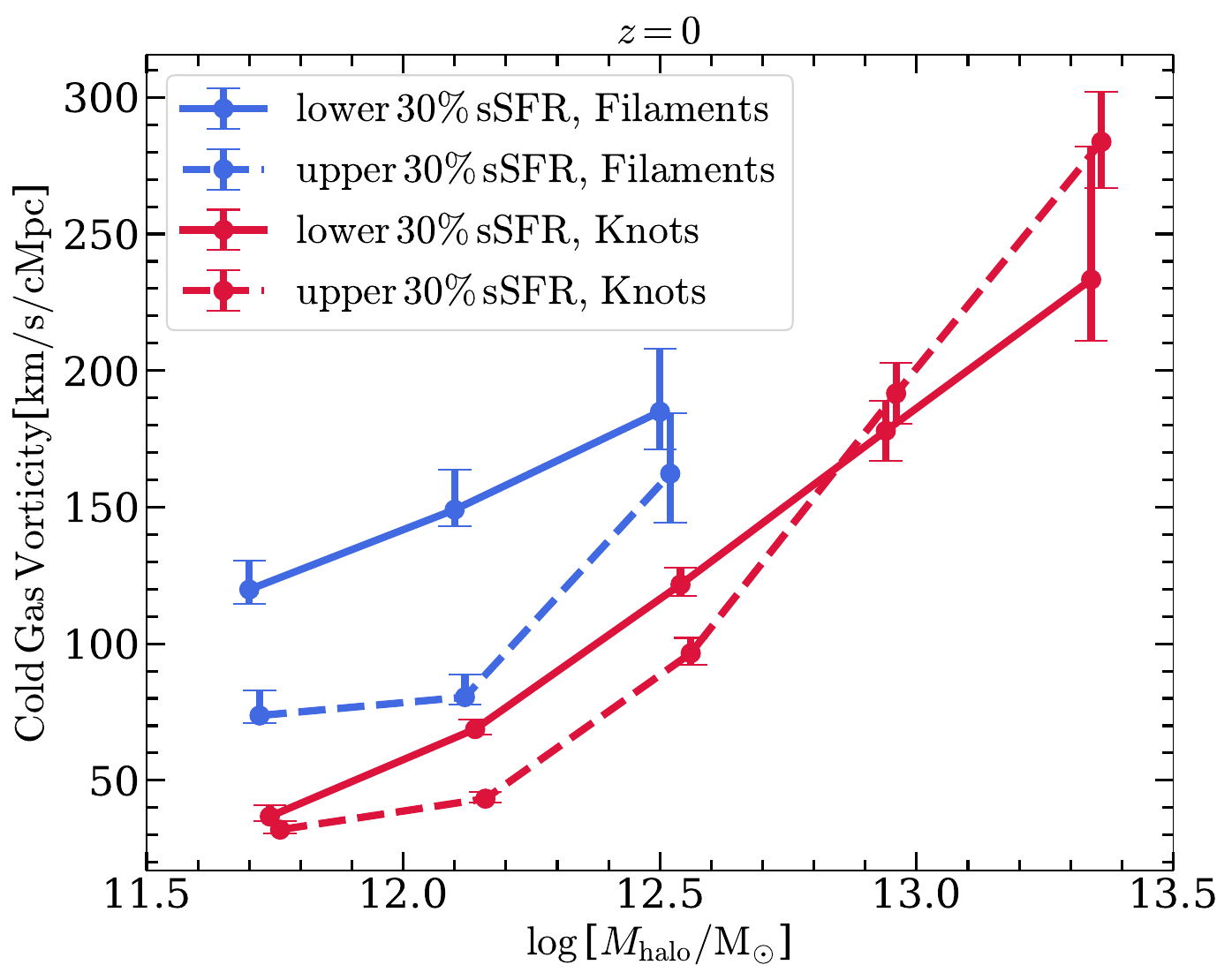}
\caption{The cold gas vorticity distributions of galaxies in filaments (blue) and knots (red). In both environments, galaxies with top (dashed lines) and bottom (solid lines) 30 per cent sSFR \textit{in each halo mass bin} are selected.}
\label{fig:Vorticity_ssfr_Binned_Env}
\end{figure}

Fig.\,\ref{fig:Vorticity_ssfr_Binned_Env} shows the cold gas vorticity distribution as a function of halo mass for galaxies living in filaments (in blue) and in knots (in red). In either case, again we select two galaxy sub-groups, i.e., the ones with  upper (dashed line) and lower (solid line) 30 per cent sSFR, within each halo mass bin. As can be seen, filaments generally host less massive halos (typically below $\log [M_{\rm halo}/{\rm M_\odot}] \sim 12.7$) than knots, this is because frequent mergers happen to knot galaxies, boosting the growth of their halo masses. At a fixed halo mass, galaxies living in filaments in general exhibit higher environmental cold-gas vorticities than those living in knots. We attribute this to the rich dynamics of the cosmic web structure (see e.g., \citealt{Pichon_2011_FilamentRichAM}). At below $\log [M_{\rm halo}/{\rm M_\odot}] \sim 12.7$ (i.e., in the first three mass bins), for galaxies in both types of environments, the sub-groups with lower (upper) 30 per cent sSFR live in an environment that has higher (lower) cold-gas vorticities, again revealing a negative modulation of the angular momentum environment to the central star formation. At the higher halo end, knot galaxies dominate the population and exhibit the opposite trend (a similar result is also presented in Fig.\,\ref{fig:ssfr_Vorticity_Binned}). We again attribute this to the correlation between the lower sSFR of galaxies and the strong AGN feedback possibly causing a lower vorticity environment (see more discussion in Section 4.1). 

\begin{figure*}
\centering 
\includegraphics[width=1\columnwidth]{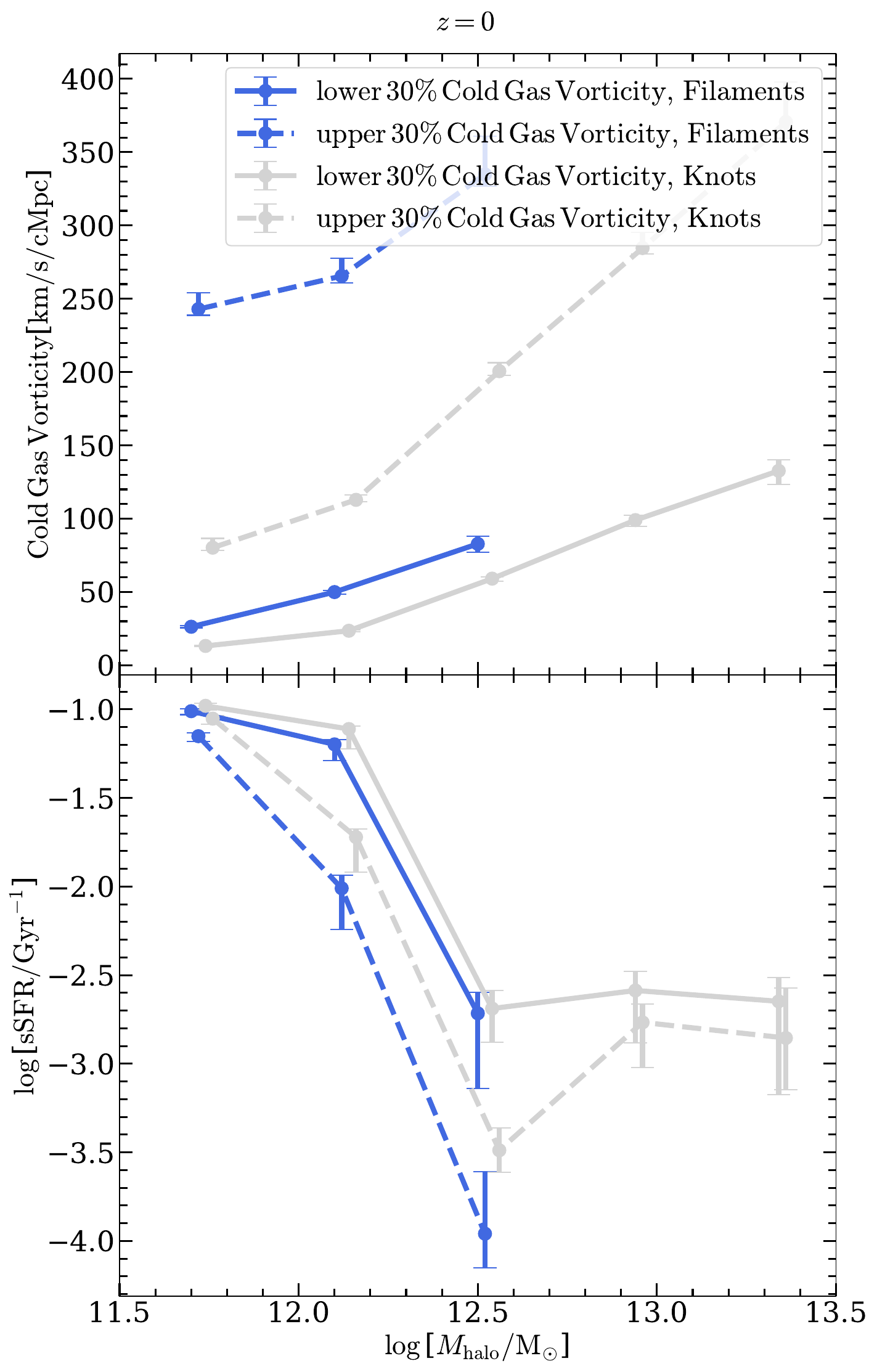}
\includegraphics[width=1\columnwidth]{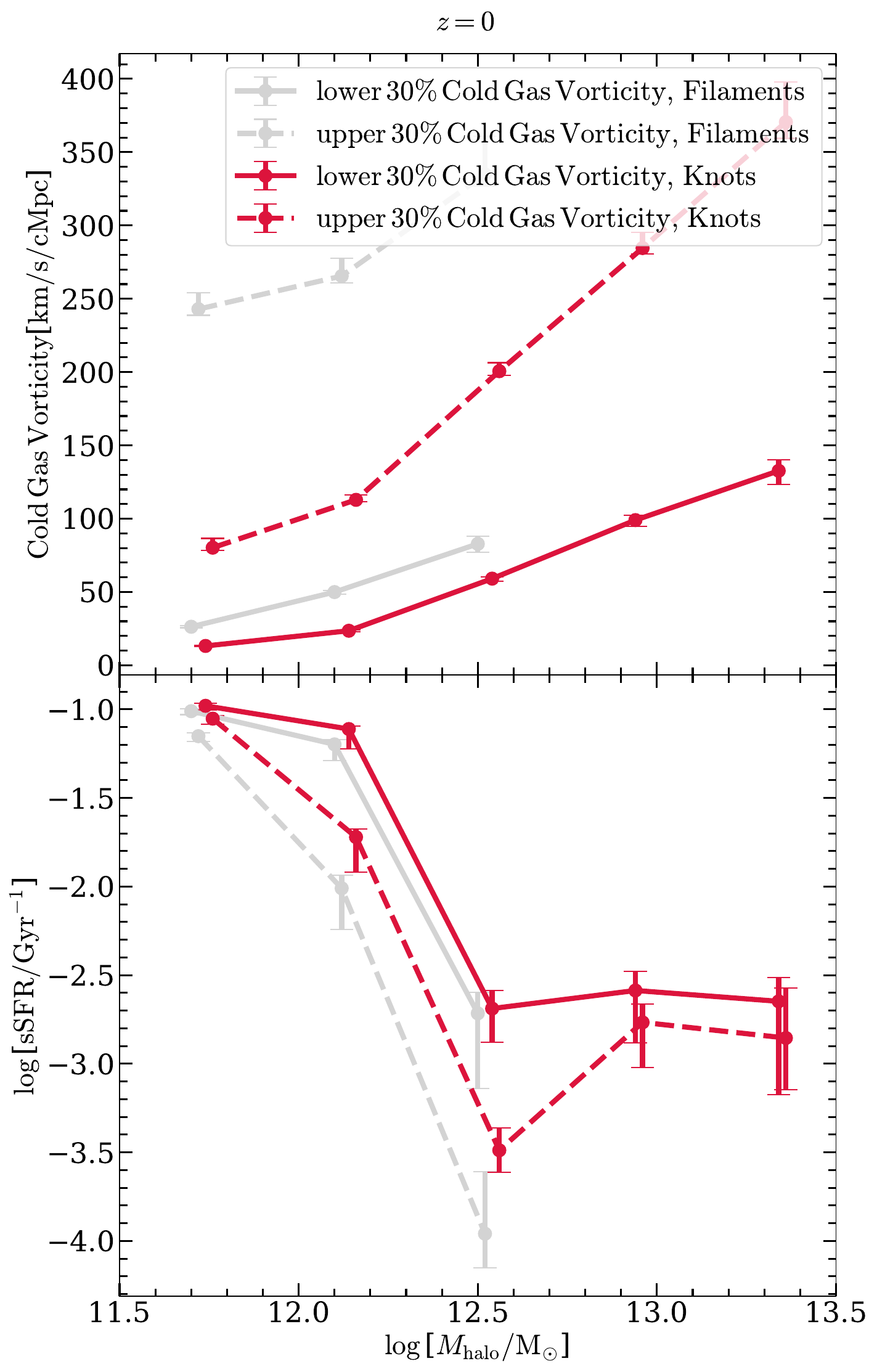} 
\caption{ Upper: Similar to Fig.\,\ref{fig:ssfr_Vorticity_Binned_SF_QG}, top and bottom panels show the galaxy cold gas vorticity and sSFR distributions respectively. Samples are also divided according to environment type, with blue lines for filament galaxies and red ones for knot galaxies. Dashed lines represent the results of galaxies with top 30 per cent cold gas vorticity \textit{in each halo mass bin} while solid lines are for galaxies with bottom 30 per cent cold gas vorticity. For visual convenience, we shade the knot (filament) data in left (right) panel. 
} 
\label{fig:ssfr_Vorticity_Binned_Env}
\end{figure*}

We then select galaxy sub-groups according to their environmental cold gas vorticities (upper and lower 30 per cent) within each halo mass bin and for a given environmental type. The median sSFR are then compared between the two subgroups at any given halo mass. The results in the filament and knot environments are shown in the left and right panels of  Fig.\,\ref{fig:ssfr_Vorticity_Binned_Env}, respectively. The negative correspondence between  higher (lower) environmental cold-gas vorticities and lower (higher) sSFR is again revealed by the dashed (solid) lines. It also can be seen that the upper (and lower) 30 per cent cold gas vorticities of the filament environment are systematically higher than that of the knot environment (comparing the left- and right-hand sides in the upper panels). This naturally explains the comparions in sSFR between galaxies from the two environmental types, in that filament galaxies have systematically lower sSFR than the knot galaxies for any given halo mass, as presented in the lower panels of the figure. 


It is worth noting that the preventative angular momentum modulation to star formation that we propose here shares the very same spirit as the one discussed in \citet{FilamentEdgeSong2021}. However we note the reader that there is a rich origin in the cold gas vorticity in a galaxy's neighborhood. For example, in edge of filament, multiple matter flows from the cosmic web structure can build up the gas vorticity (e.g., \citealt{PinchonBernardeau99, Hahn_2015_CosmicVelocityField, Laigle_2015_FilaVorticesDMhaloSpin}). As also demonstrated by our previous studies, galaxy interaction (merging or fly-by) can also be a rich source generating highly localized cold-gas streams which naturally carry the angular momentum of the incoming companions, resulting in high vorticity in the cold circumgalactic gas (as is further discussed in Section \ref{neighbour_Env_AM}). The main goal of this study is not to identify locations of high vorticity regions, and compare galaxies located in different locations in the cosmic web as has been discussed in \citet{FilamentEdgeSong2021}, but to compare galaxies with different cold-gas vorticity environments, no matter where they are located. In particular, we note that even for galaxies living in the knot environment, the cold-gas vorticity modulation still takes effect as shown in Fig.\,\ref{fig:Vorticity_ssfr_Binned_Env} and \ref{fig:ssfr_Vorticity_Binned_Env}. Through this, we point out a potential universal preventative modulation by the cold-gas vorticity on star formation activeness, as seen by the TNG100 simulation.



\section{Discussion}
\label{sec:discussion}

\subsection{In Relation to Orbital Angular Momentum of Interacting Galaxies in the Neighborhood}
\label{neighbour_Env_AM}

In \citet{Lu_et_al_2022}, the authors found a positive correlation between the orbital angular momentum of neighboring galaxies and the cold CGM spin of the central galaxy (see figure 4 and figure 9 therein), evidencing an external angular momentum modulation of the environment (from satellite galaxies) to the cold circumgalactic gas (as is also clearly evident from figure 7 and 8 in \citealt{Wang_et_al_2022}). As this is a natural consequence of gas acquisition within the context of galaxy merger and interaction, such a positive correlation shall be universally present across different simulations, although the negative correlation between the CGM spin and galaxy sSFR may or may not exist. This is indeed the case as reported by \citet{Liu_et_al_2024}, where the positive correlation was also discovered using the SIMBA simulation.

\begin{figure*}
\centering
\includegraphics[width=1\columnwidth]{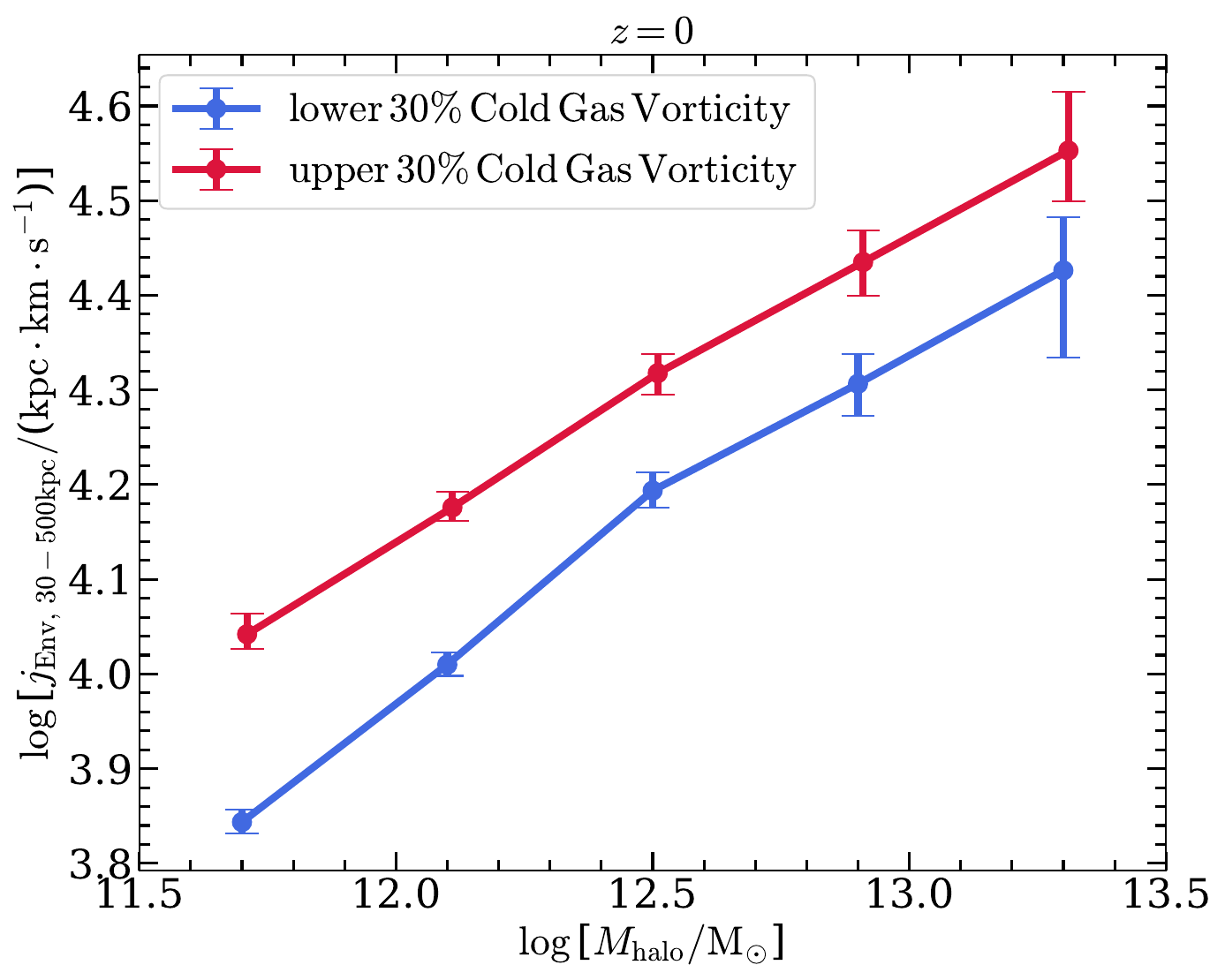} 
\includegraphics[width=1\columnwidth]{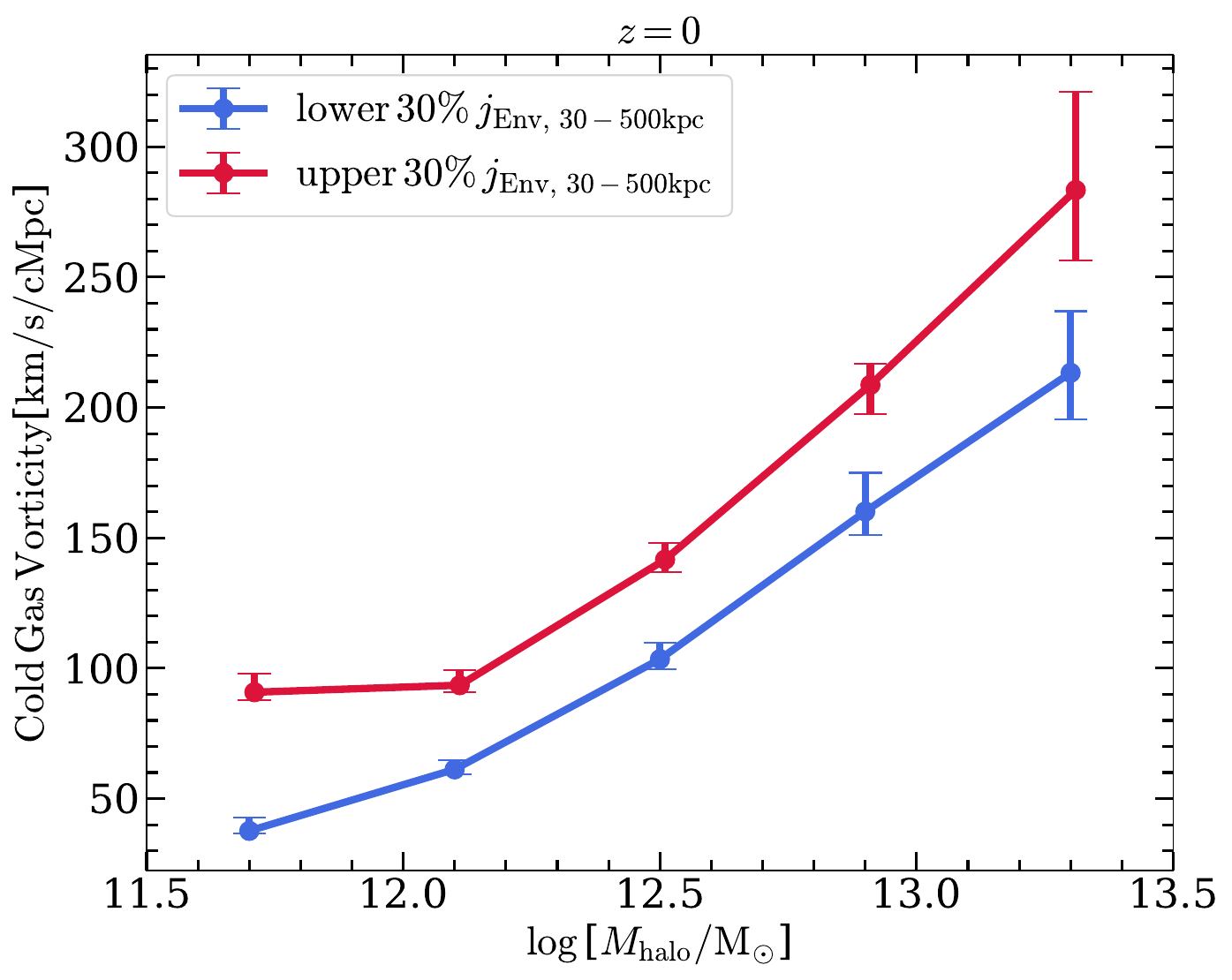}
\caption{Left: The environmental satellite orbital angular momentum distributions of galaxies living in environments with top (red) and bottom (blue) 30 per cent cold gas vorticity \textit{in each halo mass bin}. Right: The cold gas vorticity distributions of galaxies living in environments with the top (red) and bottom (blue) 30 per cent satellite orbital angular momentum \textit{in each halo mass bin}.} 
\label{fig:Env_AM_Vorticity_Binned}
\end{figure*}

Here we extend such investigation into the cosmic environment on larger scales and search for correlations between a galaxy's environmental cold gas vorticity and the orbital (specific) angular momentum from its close neighborhood. For the latter calculation, we take all satellite galaxies within a radial distance range from 30 kpc to 500 kpc from any given central galaxy. The results are presented in Fig.\,\ref{fig:Env_AM_Vorticity_Binned}. As can be seen from the top panel, galaxies that live in environments with upper (lower) 30 per cent cold-gas vorticities also exhibit higher (lower) orbital angular momenta at any given halo mass. In the bottom panel, we reverse the two properties in the plot. It can also be seen that galaxies that possess upper (lower) 30 per cent orbital angular momenta also live in environments that show higher (lower) cold-gas vorticities. As demonstrated in \citet{Wang_et_al_2022}, the kinetic motion of the cold circumgalactic gas can be significantly affected by galaxy interactions such as mergers and fly-bys (see figure 7 and 8 therein). Here we also expect that galaxy interactions may strongly affect the cold-gas vorticity field, transferring angular momentum at distances of several hundreds of kpcs and even beyond. 

It is worth noting that motions of both the cold gas on larger scales and neighboring galaxies moving along the same cosmic web structure, are tracers of the larger-scale gravitational field. Their angular momentum properties are firstly inherited from the tidal toque field established by the large-scale structure and later on affected by the rich and non-linear dynamics of matter accretion along the cosmic web. In particular, both the large-scale matter distribution and the path of the merger events through the cosmic web are a direct consequence of the cosmic evolution from an initial condition. In fact, it has been noticed by several recent studies that the angular momentum properties are already encoded by the cosmological initial condition (e.g., \citealt{Cadiou2022HighZGasTorque, Cadiou2022StellarAMInitial, Moon2024MIAMInit}).

These findings also suggest that future observational efforts towards measuring the two environmental angular momentum properties, and seeking for correlations between themselves, as well as between them and galaxy properties (such as SFR, stellar size, angular momentum, and kinematics), would provide key information for us to better understand the co-evolution between galaxies and their larger-scale environment. In particular, we propose two kinds of galaxies and their larger-scale environments as ideal test sites, i.e., superthin galaxies and low surface brightness galaxies. Theoretical studies based on cosmological simulations revealed tight correlations between their large angular momenta and large disk length-to-height ratios (\citealt{HuXuLi2024RAASuperthin}), and between their large angular momenta and their larger sizes and thus low surface brightnesses (see \citealt{Montano_et_al_2022,Montano_et_al_2024,Zhu_et_al_2023}). Both kinds of galaxies tend to live in more isolated environments, and therefore their stellar angular momenta manage to better retain angular momenta of the accreted cold gas, which are greatly modulated by the larger-scale galaxy interaction environment. We predict that these galaxies must preferably live in environments with higher cold-gas vorticities, which also exhibit more strongly coherent kinematic motions among neighboring galaxies, the CGM, and the central stellar disks across a wide distance scales (see figure 9 of \citealt{Lu_et_al_2022}).


\subsection{Vorticities in total matter, total gas and cold gas}
\label{tot_or_cold}

As already discussed, different matter components within and around galaxy halos receive different angular momenta via different modulation mechanisms. It is therefore expected to observe different angular momenta between the dark matter, gas and stellar components. The same is also true here in the case of the matter vorticity fields. In this section, we present the difference between the environmental vorticity fields of the total matter, the total gas component, and the cold gas component. The cold gas component, being a highly collisional and dissipational matter content, behaves very differently in terms of its vorticity field, from the dark matter as well as from the total gas component (which is largely dominated by the warmer and hot component within galaxy halos). As will be seen, it always exhibits the highest vorticities in the same region, due to its unique, highly collisional and dissipational nature.   

\begin{figure*}
\centering
\includegraphics[width=2\columnwidth]{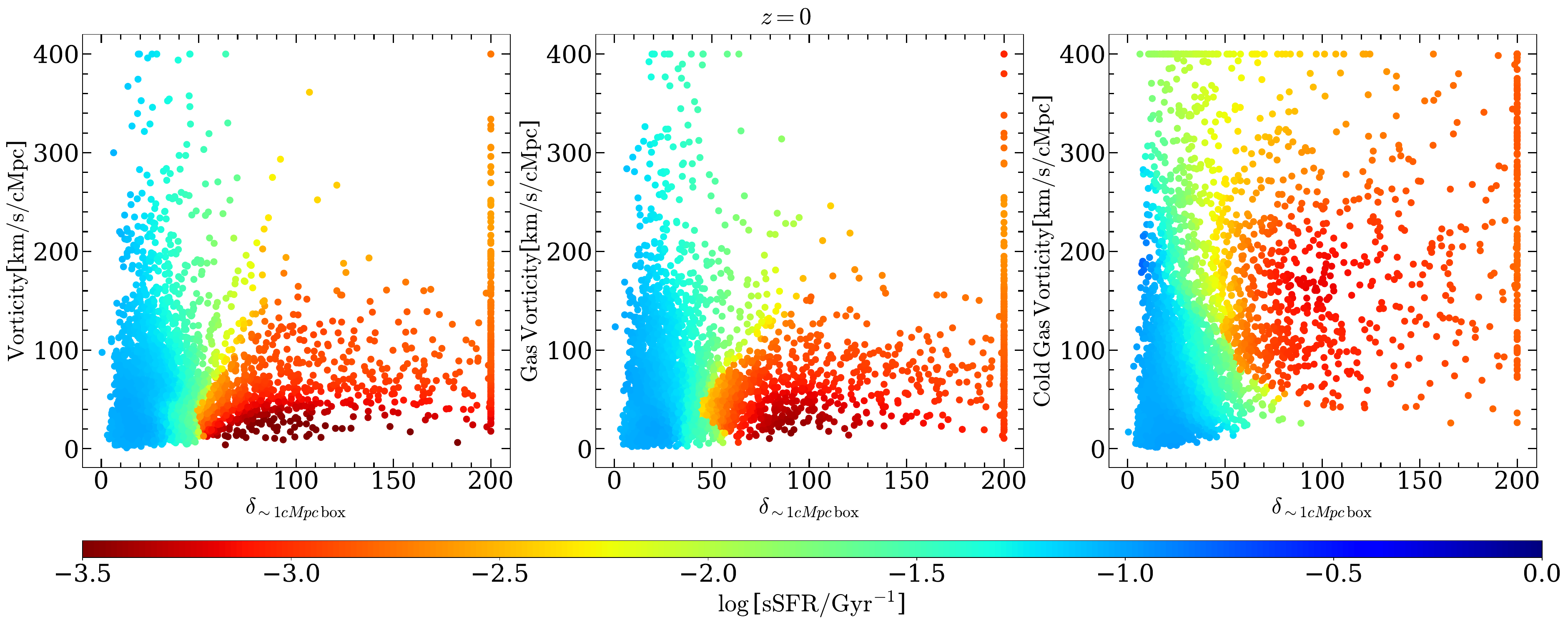}
\caption{Distributions of grid cells (with total stellar mass from {\it central} galaxies larger than $10^{10} {\rm M_{\odot}}$) on the over-density -- vorticity phase space, color-coded by the {\it cell} specific SFR at $z=0$. From left to right, the $y$-axis denotes the vorticity of the total matter, total gas, and the cold gas, respectively. We note that there is very small fraction of data points located far away beyond over-density of 200, and vorticity of 400. Therefore these threshold values are re-assigned to these points. } 
\label{fig:Vorticity_Overdensity}
\end{figure*}

This can be clearly seen from Fig.\,\ref{fig:Vorticity_Example}, where the magnitudes of the cold-gas vorticity fields are significantly larger than those of the total matter vorticity fields. The same trend is also evident in Fig.\,\ref{fig:Vorticity_Overdensity}, where we plot the field strengths of three types of vorticities, i.e., the total matter (left-most), the total gas (middle), and the cold gas (right-most), as a function of the field over-density. In this figure, only grid cells with a total stellar masses (from {\it central} galaxies) larger than $10^{10} \mathrm{M_{\odot}}$ are used. Each panel is also color-coded by the grid sSFR (averaged within a region of $\sim$1\,Mpc). As the matter over-density increases (horizontally from left to right), the region average sSFR decreases - a result widely confirmed by previous studies (e.g., \citealt{Winkel2021_cosmicwebdensitySFR, Galarraga-Espinosa2023_filamentDensitySFR, Hasan2023_CosmicWebQuench, Hasan2024_FilamentQuenching}). It can also been seen that, as the cold-gas vorticity increases (vertically from bottom to top), the sSFR also decrease significantly (the right-most panel, and also presented in Section \ref{sec:general_sample}). This trend, however, only exists for the cold-gas vorticity, but neither in the case of total matter vorticity, nor the total gas vorticity. As discussed already, this is a result of the uniqueness of cold gas, which can cool efficiently while gravitationally influenced by the cosmic web structure, forming highly localized vortical flow, in comparison to its hotter-gas or dark matter counterparts.

\subsection{Dependence on simulation subgrid recipes}

It is worth noting that the preventative modulation between the angular momentum environment and galaxy star formation as clearly evident in the TNG100 simulation might not be universally seen in other cosmological simulations which take on different subgrid models to treat feedback. For example, a negative correlation between the CGM spin and the sSFR of TNG100 galaxies as reported by \citet[see figure 5 therein]{Wang_et_al_2022} and \citet[see figure 3 therein]{Lu_et_al_2022} is not present for the SIMBA galaxies as reported by \citet[see figure 2 therein]{Liu_et_al_2024}. One key reason is that for a causal connection to establish between the large-scale cold-gas angular momentum and the central star-formation activity, there shall be a crucial mechanism that can effectively transfer the CGM angular moment to the star-forming ISM through mixing, which strongly depends on the detailed subgrid models for the ISM feedback. A recent study by \citet{Yang_et_al_2024} compared galaxy specific angular momenta between the Apostle (see \citealt{Sawala_et_al_2016,Fattahi_et_al_2016}, a family of the EAGLE simulation suite, \citealt{Schaye2015Eagle}) and Auriga (\citealt{Grand_et_al_2017}, a family of the TNG suite) simulations and clearly demonstrated so. Due to different subgrid implementations, the Auriga galaxies manage to maintain a high fraction of gas going through a recycled fountain flow which acquires angular momentum from CGM through mixing; while the Apostle galaxies only possess a much smaller fraction of gas undergoing the recycled fountain process and thus having experienced negligible angular momentum acquisition. It is therefore worth examining such signals using different cosmological simulations.

\section{Conclusions}
\label{sec:conclusion}
In this study, we take $z=0$ galaxies from the TNG100 simulation and investigate the influence of the large-scale angular momentum environment as depicted by a cold-gas vorticity, on a galaxy's star formation activeness depicted by sSFR. This vorticity calculated for gas with $T_{\rm gas} < 2\times 10^4 \mathrm{K}$ and on scales of $\sim$1\,Mpc can well describe the angular motion of the ambient cold gas. We find crucial evidence for the connection between the cold gas spin/vorticity and star formation activeness. We present our main conclusions as follows:

\begin{itemize}
\item At any given halo mass, galaxies that live in a higher cold-gas vorticity environment are generally less actively star forming. Specifically, at a fixed halo mass scale of $10^{12}-10^{13}\,\mathrm{M_{\odot}}$, the median sSFR of galaxies living in an environment with the top 30\% cold-gas vorticity is $\sim 0.5$ dex below that of galaxies living in an environment with the bottom 30\% cold-gas vorticity (see Section \ref{ssfr_vorticity}).

\item It is observed among both star-forming and quenched galaxy populations that at any given halo mass, galaxies living in higher cold-gas vorticity environments are less actively star forming  (Section \ref{starforming_quenched}). 

\item 
At any fixed halo mass scale, the cold-gas vorticities in filaments are generally higher than that in knots, naturally explaining lower sSFRs of the filament galaxies than of the knot galaxies (Section \ref{dep_Env_typ}). 

\item The cold-gas vorticity is highly connected to the orbital angular momentum of neighboring galaxies, indicating their common origins and potential angular momentum inheritance/modulation from the latter to the former (Section \ref{neighbour_Env_AM}). 

\item The negative modulation by the ambient vorticity field to galaxy star formation is only significantly observed for the cold gas, but neither for the total matter nor for the total gas component, indicating the unique role of the ambient cold gas (Section \ref{tot_or_cold}).

\end{itemize}

We note the reader that the evidence that we have shown here for higher cold-gas vorticity preventing efficient gas infall, resulting in lower sSFR, is obtained using the simulated galaxies at $z=0$. While at higher redshifts, the situation might be different, which is not investigated in this study. However, it is worth noting that \citet{Kocjan2024HiZColdAM} used ultra-high resolution hydrodynamical simulations of galaxy formation to investigate relations between gas accretion and star formation at $z=2$. They found that apart from shock-heating and outflow of hot gas, the cold gas with lower specific angular momentum also contributes to the delay of star formation, as it is trapped in a turbulent structure due to higher levels of disordered motions and thus only slowly accreted to form stars, while the higher-angular momentum cold gas is rather fast accreted and not a cause of the delay of star formation. This work highlights the importance of investigating the combined effect of gas specific angular momentum and gas accretion time, which will be carried out in future studies. 

Through this study we emphasize that the angular/vortical motion on large scales, as manifested through the vorticity field, in particular that associated with the cold gas, can be a crucial element shaping the star-forming and quenching status of a galaxy, and thus an important feature, among others, to depict a galaxy’s larger-scale environment, when studying the co-evolution between the galaxies and their large-scale environments.


\section*{Acknowledgements}
We acknowledge Profs. Volker Springel, Simon White, Cheng Li and Yong Shi for their constructive suggestions that help to better deliver the key idea of the study. We also thank the anonymous referee for constructive comments which help significantly to improve the quality of the paper. This work is supported by the National Key Research Development Program of China (Grant No. 2022YFA1602903) and the National Natural Science Foundation of China (Grant No. 12433003). This work acknowledges the high-performance computing clusters at the Department of Astronomy, Tsinghua University.

%




\bibliography{sample631}{}
\bibliographystyle{aasjournal}



\end{document}